\documentclass[prb, twocolumn, superscriptaddress,showpacs,aps]{revtex4-1}

\usepackage{amsmath}
\usepackage{amstext}
\usepackage{graphicx}
\usepackage{rotating}
\usepackage{tabularx}
\usepackage{color}
\usepackage[english]{babel}
\usepackage{times}

\def\ket#1{\left|#1\right\rangle}

\def\avg#1{\left\langle#1\right\rangle}

\begin{document}

\title{1/f Flux Noise in low-T$_c$ SQUIDs due to Superparamagnetic Phase Transitions in Defect Clusters}

\author{Amrit De}
\email[E-mail:]{amritde@gmail.com}
\affiliation{Department of Electrical Engineering, University of California - Riverside, CA 92521}

\date{\today}

\begin{abstract}
It is shown here that $1/f^\alpha$ flux noise in conventional low-T$_c$ SQUIDs is a result of low temperature
superparamagnetic phase transitions in small clusters of strongly correlated color center defects.
The spins in each cluster interact via long-range ferromagnetic interactions. Due to its small size, the cluster behaves like a \emph{random-telegraphic} macro-spin when transitioning to the superparamagnetic phase.
This results in $1/f^{\alpha}$ noise when ensemble averaged over a random distribution of clusters.
This model is self-consistent and explains all related experimental results which includes $\alpha\sim 0.8$ independent of system-size. The experimental flux-inductance-noise spectrum is explained through three-point correlation calculations and time reversal symmetry breaking arguments.
Also, unlike the flux noise, it is shown why the second-spectrum inductance noise is inherently temperature dependent due to the fluctuation-dissipation theorem.
A correlation-function calculation methodology using Ising-Glauber dynamics was key for obtaining these results.
\end{abstract}

\maketitle

\section{Introduction}\label{sec.intro}

\begin{figure}\center
\includegraphics[width=0.9\columnwidth]{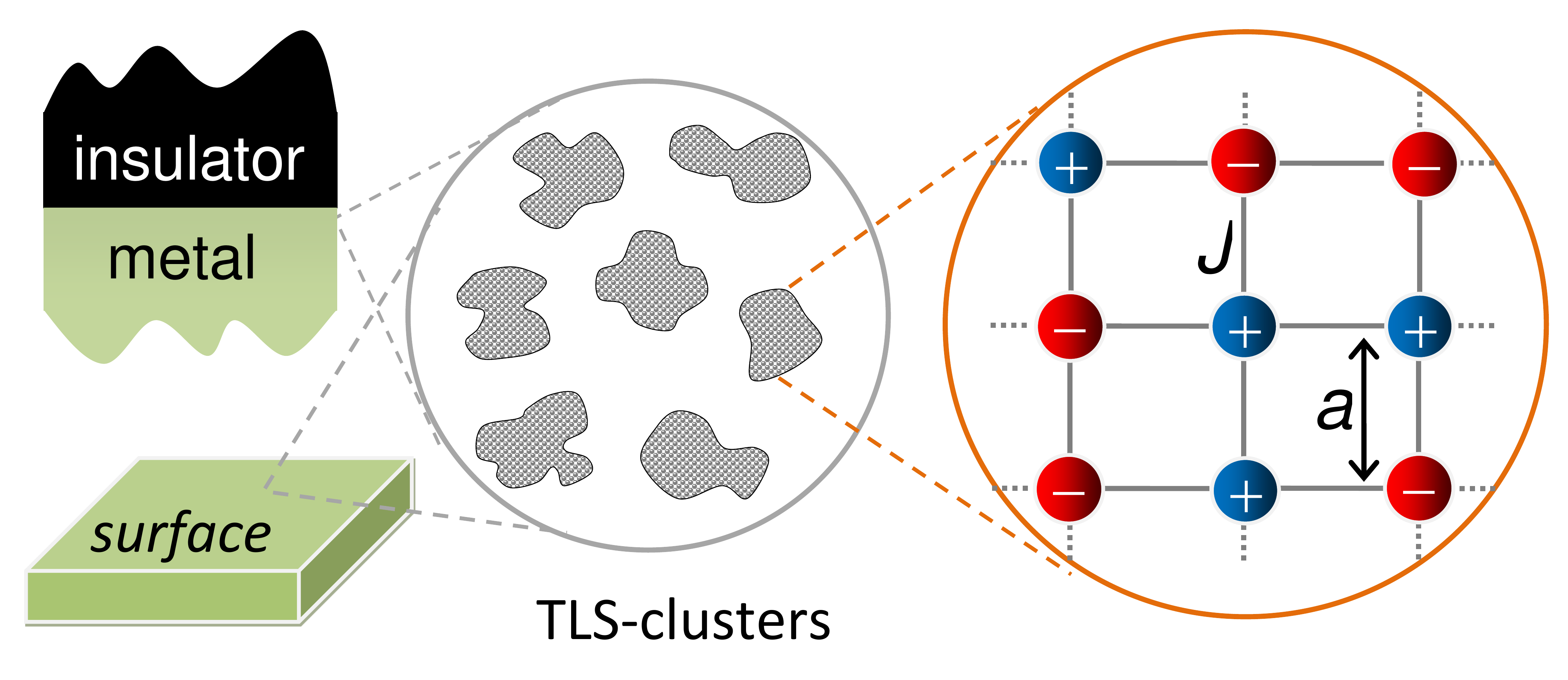}
\caption{$1/f^{\alpha}$ noise model consisting of interacting and fluctuating TLSs in a cluster. The clusters are assumed to form at the metal-insulator interface or on the surface. They are sufficiently far apart so that only spins within a single cluster interact. Number of TLSs within a cluster and the lattice constant $a$ vary.}
\label{fig:scheme}
\end{figure}

Ever since its first measurement in the 1920's\cite{Schottky1926}, flicker noise or $1/f$ noise has been seen in a wide variety of solid state systems\cite{Dutta1981,Weissman1988,Paladino2014}. Examples include spin glasses\cite{Massey1997,Wu2005,Weissman1993}, Coulomb glasses\cite{Shtengel2003,Shumilin2014},metal films\cite{Voss1976}, metal-insulator tunnel junctions\cite{Manatese1985,Garfunkel1988,Rogers1984}, various semiconductor devices\cite{Fleetwood1994} such as
such as field-effect transistors(FETs)\cite{Uren1985,Lai2014}, core-shell nanowire FETs\cite{Persson2013}, GaN/AlGaN heterostructures\cite{Balandin1999} and more recently in Graphene devices\cite{Kaverzin2012,Balandin2013}. Though there is no common physical underlying mechanism that gives rise to all these different manifestations\cite{Dutta1981}, it has been argued by Bak that $1/f$ noise will occur in barely stable dynamical systems with extended spatial degrees of freedom\cite{Bak1987} which evolve into self organized critical structures.

In many solid state systems the presence of parasitic two-level-systems(TLSs), possibly due to the presence of defects, generates random-telegraphic-noise(RTN)\cite{Kogan.book,Galperin2006}. Typically if there are a large number of fluctuating TLSs then a log normal distribution of their switching rates gives rise to a $1/f$ power spectrum within some frequency range.
For solid state quantum computing, $1/f$ noise is a major problem as it is a significant source of decoherence\cite{Paladino2014,Bellomo2008,Dajka2008,Burkard2009,Zhou2010}. In semiconductor quantum dots(QDs), RTN is observed when  electrons randomly tunnel back and forth\cite{Vandersypen2004,MacLean2007,Taubert2008}.

Currently low-$T_c$ superconducting quantum interference devices (SQUIDs) are at the forefront of quantum computing. SQUID based qubits are promising as they can replicate properties of natural qubits (such as electron and nuclear spins) using macroscopic devices\cite{Wendin2007}. However, the practical implementation of a scalable quantum computer based on charge-, flux-, phase- or transmon qubits is severely impeded by the presence of $1/f$ charge noise or $1/f$ magnetic noise. This limits the phase coherence of SQUID based qubits\cite{Paladino2014}.

High-$T_c$ SQUIDs, such as those made from YBCO, have high $1/f$ flux noise due to vortex motions
\cite{Koelle1999}. A key feature of this type of flux noise is its dependence on the on the rate of change of temperature\cite{Monaco2006,Gordeeva2010,Weir2013}. Whereas low-$T_c$ SQUIDs generally do not have 1/f flux noise problems from vortex motion.

This paper focuses on various puzzling features of $1/f$ magnetization noise in low-$T_c$ SQUIDs. In the case of flux based qubits and phase qubits, experiments have revealed that the magnetic flux noise has a $1/f^{\alpha}$ power spectrum \cite{Yoshihara2006,Bialczak2007}. Although, this type of magnetic flux noise was first observed in SQUIDs in the 80's\cite{Koch1983,Wellstood1987}, its origins were not fully explained. The interest in this subject has however been revived because of the recent activity in quantum computing. A better understanding of the microscopic origins of this noise could lead to better Joshephson tunnel junction designs\cite{Oh2006} and possible elimination of a source of contamination in SQUIDs, such as by surface treatment\cite{Kumar2016pra}.

Magnetic noise in SQUIDs have several key features. Amongst them is that the flux noise is only weakly dependent on geometry. Recent measurements on qubits with different geometries indicate that the flux noise scales as $l/w$, in the limit $w/l <<1$ (where $l$ is the length and $w$ is the width of the superconducting wires)\cite{Lanting2009}. This along with recent experiments by Sendelbach {\it et al} \cite{Sendelbach2008} suggests that flux noise arises from unpaired surface spins which reside at the superconductor-insulator interface in thin-film SQUIDs. The estimated areal spin density from the paramagnetic susceptibilities, for superconductor-insulator\cite{Sendelbach2008} interfaces (and metal-insulator\cite{Bluhm2009} interfaces) is about $5\times10^{17}~m^{-2}$. As a result of this high spin density, the coherent magnetization of the spins results in a large flux coupling to the SQUID.


Another key feature of this flux noise is that it is only weakly dependent on parameters such as temperature, choice of the superconducting material and the area of the SQUID\cite{Koch1983,McDermott2009}. In addition to the flux noise, the inductance noise was also measured in the experiments of Ref.[\onlinecite{Sendelbach2008}]. It was seen that inductance noise, which scales as $1/f^{\alpha}$, decreases with increasing temperature and $\alpha$ itself is temperature dependent ($0<\alpha(T)<1$) \cite{Wellstood2011}.

This deduced areal density is consistent with some of the theoretical models such as in terms of metal induced gap states that arise due to the potential disorder at the metal-insulator interface\cite{Choi2009}.
Several models have been suggested including non-interacting electron hopping between traps with different spin orientations
and a $1/f$ distribution of trap energies\cite{Koch2007}; 
a dangling bond model\cite{deSousa2007} and  interacting fractal spin clusters with varying number of spins to obtain a $1/f$ distribution of relaxation rates\cite{Kechedzhi2011} and a flux-vector model\cite{LaForest2015prb}.
Sometime ago Kozub\cite{Kozub1996} proposed a model where the amplitude of $1/f$ noise scaled as $1/T$, however this model did not consider spin-spin interactions.

Experimental evidence on the other hand suggests that these surface spins are strongly interacting and that there is a net spin polarization \cite{Sendelbach2008}. It is seen that the $1/f$ inductance noise is highly correlated with the usual $1/f$ flux noise. Their cross-correlation is inversely proportional to the temperature and is about the order of unity roughly below T ~100mK. Now inductance is even under time inversion whereas flux is odd under time inversion. This implies that their three-point cross-correlation function must be zero unless time inversion symmetry is broken. This is only possible only by the appearance of long range magnetic ordering, unless an external magnetic field is applied. The mechanism producing both the flux noise and inductance noise is expected to be the same.

It has been suggested that the spins at the superconductor-insulator interface interact with each other via the Ruderman–Kittel–Kasuya–Yosida (RKKY) mechanism\cite{Faoro2008} which is responsible for the spin polarization reported in experiments\cite{Sendelbach2008}. The RKKY interaction can give rise to unusual magnetic ordering in SQUIDs as the magnetic ordering can oscillate between being ferromagnetic to being antiferromagnetic as a function of distance. Antiferromagnetic RKKY interactions can give rise to a spin glass type phase and $1/f$ noise related to the magnetic fluctuations and low temperature kinetics \cite{Massey1997,Wu2005,Weissman1993,Shklovskii1980,Shklovskii2003,De2015noise}.


However Monte-Carlo simulations\cite{Chen2010}, ruled out the formation of a spin glass phase to explain magnetization noise in SQUIDs using an Ising model with random nearest neighbor interactions. The model was shown to qualitatively reproduce  temperature dependent inductance noise features, but did not show the time-reversal-symmetry-breaking cross-correlations between inductance and flux noise. This is expected in spin glasses due to zero net magnetic moment, which leads to a vanishing three-point cross-correlations between magnetization and susceptibility. In Ref.\onlinecite{De2014PRL2}, we also found that \emph{antiferromagnetic} RKKY interactions for small spin clusters, resulted in negligibly small three-point correlations.

In a related development, a few years ago surface ferromagnetism (SFM) was reported in thin-films and nanoparticles of a number of otherwise insulating metallic oxides\cite{Sundaresan_prb2006} (including Al$_2$O$_3$) where the materials were not doped with any magnetic impurities. Further recent investigations attribute this room temperature SFM  in Al$_2$O$_3$ nanoparticles\cite{Yang_JPC2011} to Farbe${+}$(F$^+$)-center where it was found that amorphous Al$_2$O$_3$ is more likely to host the number of F$^+$-centers to cross the magnetic percolation threshold than the crystalline variant. The origin of SFM in these otherwise non-magnetic metal oxides is itself somewhat controversial\cite{Keating2009} where a number of different exchange coupling mechanisms have been proposed \cite{Venkatesan2004,Han_prb2009,Chang_PRB2012}.

SFM and the SQUID geometry provides some important clues on the microscopic origins of the $1/f$ flux noise.
Typically low-$T_c$ dc-SQUIDs have an amorphous Al$_{2}$O$_{3}$ insulating layer deposited on the surface of a metal(commonly Nb\cite{Sendelbach2008} or Al\cite{Bialczak2007}). Al$_{2}$O$_{3}$ is likely to cluster on the surface before filling in and forming a homogeneous layer due to its higher binding energy which could lead to the Volmer-Weber growth mode. The lattice mismatch between the insulator and the metal could also lead to the formation of clusters. Near the metal surface, the clusters can host a number of point defects in the form of O vacancies that can capture one electron -- Farbe${+}$(F$^+$)-center, or two (F-center).
Surface absorption of $O_2$\cite{Wang2015prl,Kumar2016pra}, intrinsic vacancies\cite{Lee2014PRL} and even hydrogen\cite{Wang2017arxiv} are among some of the suggested origins of the magnetic moments responsible for the flux noise.

If the magnetic moments are at the SQUID's metal-insulator interface, then they can interact via the RKKY interaction\cite{Faoro2008}.
Because of the proximity to the metal, these local magnetic moments can spin polarize the metal's conduction band electrons which can lead to an RKKY type long range interaction. This leads to competing interaction mechanisms.
However if these parasitic magnetic moments are on the surface then other long range ferromagnetic interactions come into play\cite{Kumar2016pra}. Recently it has been suggested that the TLS interact via phonon modes\cite{Lisenfeld2016}. In order for this paper's results to be applicable and to explain the flux-inductance cross spectrum noise, the long range phonon mediated TLS-TLS interactions need to be \emph{ferromagnetic}.


A temperature dependent spin-cluster model with ferromagnetic RKKY interactions was proposed recently to explain various puzzling features of $1/f$ magnetization noise in SQUIDs\cite{De2014PRL2}. This spin-cluster model explains various experimental results self consistently and is representative of a disordered system at the SQUID's metal-insulator interface (see fig.\ref{fig:scheme}).
The results are nearly identical with other types \emph{long-range} ferromagnetic interactions which applies to the $O_2$ surface absorption picture $\cite{Wang2015prl,Kumar2016pra}$.
However {short-range} ferromagnetic interactions lead to flux noise that varied considerably more with temperature due to weaker correlations.
The current paper builds on this previous work\cite{De2014PRL2}.

A key relation between the noise exponent $\alpha$ and superparamagnetic phase transitions (SPTs) is uncovered here while trying to explain the measured flux-inductance noise cross correlations. Sharp looking SPTs are shown to occur, for a single cluster, at temperatures much lower than the Curie temperature.
Even the smallest temperature fluctuations will then result in \emph{random telegraphic} magnetic noise.
This is because each cluster behaves like a fluctuating macro-spin because of SPTs.
Using three-point correlation function calculations, it is shown how this relates to the experimentally observed flux noise and inductance noise cross-spectrum noise.

In general, SPTs are well known to occur in single domain nano-magnets\cite{Dormann1988,Knobel2008,Petracic2010,Skumryev2003,Eisenmenger2003}.
Similarly here, because of the small cluster size, the magnetic anisotropy energy per particle can become comparable to the thermal energy, which leads to SPTs. Hence the ferromagnetic to superparamagnetic phase boundaries are very sharp looking even though each cluster size is small. The long-range ferromagnetic RKKY interactions makes the cluster very strongly-correlated.

Additionally, the SPTs are also accompanied by the expected smooth superparamagnetic to paramagnetic crossovers at the higher Curie temperatures. An experimentally observation of this would validate this model.
Whereas experimentally, the sharp SPTs would be difficult to observe directly for a single defect cluster.
However, these SPTs will lead to the observable $1/f^{\alpha}$ magnetization noise as a result of the cluster ensemble.
When considering multiple random clusters (with a spread in the critical temperatures) -- it is shown that $1/f^\alpha$ noise occurs in the same temperature range where the SPTs occur. $\alpha(T)\sim 0.8$ over a range of temperatures and $\alpha$ drops off as the system becomes superparamagnetic.
Experimental evidence for a dynamical paramagnetic environment has also recently emerged from the frequency-asymmetric 1/f flux noise at very low temperatures\cite{Quintana2017}.


The model used in this paper is fully self-consistent.
Multiple spin clusters are considered with a normal random distribution of lattice constants, which is representative of defects.
It is shown here that $1/f^{\alpha}$ noise (where $\alpha\sim0.8$) arises naturally from the just following two -- long range ferromagnetic spin-spin interactions and multiple spin-clusters with random spin-spacing. Both of these features must be there for the $1/f^{\alpha}$ flux noise.
The spin-flip rates here are determined by Ising-Glauber dynamics.
There is no \emph{prior} assumption on a log-normal distribution of fluctuation rates to get $1/f$ noise.
Such a heuristic assumption typically leads to $\alpha\sim1$ instead of the $\alpha\sim0.8$, obtained here and in experiments\cite{Sendelbach2008,Kumar2016pra}.
Furthermore the flux noise is shown to be independent of system size -- similar to experiments.

Next, the inductance noise spectra, which is also the second spectrum (or the noise of the noise), is also extensively discussed here.
It is not obvious as to why, the experimentally measured second spectra shows a huge temperature dependence while the first spectrum does not even though they have the same underlaying noise microscopics.
Here it is analytically shown why the measured inductance noise inherently has a huge $T^{-2}$ temperature dependence, even though the flux noise (first spectrum) does not have any such dependence. This dependence arises from the fluctuation-dissipation theorem.
Analytical expressions are provided for the 4-point correlation power spectra.


A new correlation-function calculation technique is key to all of these results.
The suggested method systematically extracts any $n^{th}$ order correlation function for $N$ interacting Ising spins, within the framework of Ising-Glauber dynamics. Time-correlations, spatial-correlations, interactions and temperature are all taken into account.
Detailed discussions are presented in this paper along with some simple examples. Overall this method is well suited for numerics as well as for analytics in smaller systems
and is inspired by the quasi-Hamiltonian open quantum systems formalism\cite{Cheng2008,Joynt2011,Zhou2010,Zhou2010b,De2011,Zhou2012,De2013pra}.

This paper is organized as follows. In section-\ref{sec:model} the model and the technique for calculating the correlation functions is discussed. The flux noise is discussed in sec.\ref{sec:flux} along with the noise exponent in sec.-\ref{sec:alf}
The most important magnetic phase transition results and its relation to $1/f^{\alpha}$ noise are presented in sec.-\ref{sec:cross}.
The flux-inductance noise cross-spectrum, three point correlations are discussed here.
In sec.-\ref{sec:MC}, the higher Curie temperature pseudo phase-transitions are compared against the noise exponent.
Monte Carlo simulations are used only in this section.
This is followed by the inductance noise calculations along with various analytical expressions are presented in sec.-\ref{sec:Ind}.
The summary is followed by the Appendix-A where a two-spin example is worked out. The Gaussian approximation for four-point correlations is discussed in Appendix-B.

\section{Method: Correlation Functions from Ising-Glauber Dynamics}\label{sec:model}

Fluctuating two level systems can be treated as Ising spins which flip randomly in time. Their stochastic dynamics is therefore governed by the master equation \cite{VanKampen.book},
\begin{equation}
\frac{d{\mathbf{W}}(t)}{dt}=\mathbf{VW{\rm\it(t)}}
\label{ME}
\end{equation}
where $\mathbf{V}$ is a matrix of transition rates (such that the sum of each of its columns is zero) and $\mathbf{W}$ is the flipping probability matrix for the TLS. For $N$ TLS, $\mathbf{V}$ and $\mathbf{W}$ are $2^N\times2^N$ matrices.

For correlated spin fluctuations, the system's overall temporal dynamics is also governed by the master equation Eq.\ref{ME}. The Ising-Glauber model (also known as the kinetic Ising model) can be used to treat the non-equilibrium dynamics for fluctuating spins\cite{Glauber1963,Ozeki1997}. Single-site Glauber dynamics requires that
a single spin is flipped at a given site, and that the new configuration agrees with the old one everywhere except where the spin was flipped. This is a Markov process where the new distribution of spins depends only on the current spin configuration. And for Glauber dynamics, the conditional probability for a spin to flip is determined by the Boltzmann factor. The matrix-elements of $\mathbf{V}$ for correlated spin flips are therefore
\begin{equation}
\small{
{\mathbf{V}}(\mathbf{s}\rightarrow \mathbf{s}^{\prime })=\left\{
\begin{array}{ll}
\displaystyle\frac{\gamma e^{-\beta H(\mathbf{s^{\prime }})}}{e^{-\beta H(%
\mathbf{s})}+e^{-\beta H(\mathbf{s^{\prime }})}} &
\mbox{for
$\mathbf{s}\neq\mathbf{s}'$ and} \\
& \mbox{$\displaystyle\sum_i(1-s_is_i')=2$} \\
-\displaystyle\sum_{\mathbf{s}\neq \mathbf{s}^{\prime }}V(\mathbf{s}%
\rightarrow \mathbf{s}^{\prime }) & \mbox{for $\mathbf{s}=\mathbf{s}'$}%
\end{array}%
\right.}   \label{Glaub}
\end{equation}%
Here, $\mathbf{s}'$ is a vector denoting the present spin configuration of the lattice, $\mathbf{s}$ denotes the spin configuration of the lattice at an earlier instance of time and $\gamma$ is the relaxation rate of the spin that is flipped. The non negative off-diagonal matrix elements in Eq.\ref{Glaub} satisfy the detailed balance condition and the diagonal terms is the just negative sum of the off-diagonal column elements so that all column elements sum up to zero, which ensures the conservation of probability. The systems temporal dynamics is then governed by the flipping probability matrix, which is $\mathbf{W}=\exp(-\mathbf{V}dt)$. The eigenvalues of $\mathbf{V}$ are either zero, which corresponds to the equilibrium distribution, or are real-negative which also eventually tend to the equilibrium distribution as time $t\rightarrow\infty$ \cite{VanKampen.book}.

The overall system Hamiltonian in the Boltzmann factor in Eq.\ref{Glaub} is
\begin{equation}
H(\mathbf{s}) = -\frac{1}{2}\sum_{i,j}J_{ij}s_is_j - B\sum_i s_i
\label{H_ising}
\end{equation}
where $B$ is the magnetic field and $J_{ij}$ is the spin-spin interaction between the $i^{th}$ and $j^{th}$ spins.
In this paper two types of $J_{ij}$s are considered.
For the first type, it is assumed that within a single cluster, the spins interact via an oscillatory RKKY-like form:
\begin{eqnarray}
J_{ij}=J_{o}\frac{[k_FR_{ij}\cos (k_FR_{ij})-\sin (k_FR_{ij})]}{(k_FR_{ij})^{4}}
\label{J_RKKY}
\end{eqnarray}%
where $J_o$ is assumed to be ferromagnetic.
Here $R_{ij}$ is the separation between two spins (on a lattice of lattice constant $a$), $k_F$ is a Fermi wavevector type parameter. For the calculations here $J_o\approx 10^{11}~Hz/\hbar$ is taken as a fitting parameter independent of $k_F$.
A short range ferromagnetic nearest neighbor interaction (NNI) is also considered where $J_{ij}\propto{1/R_{ij}}$.


In general, for $N$ spins (either interacting or non-interacting), any $n^{th}$ order correlation function between arbitrary spins can be \emph{exactly} calculated as follows
\begin{eqnarray}
\langle s_{\ell}(t_{1})s_{j}(t_{2})...s_{\kappa}(t_{n})\rangle = ~~~~~~~~~~~~~~~~~~~~\\
\langle {\mathbf{f}}|\sigma_z^{(\kappa)}\mathbf{W}(t_n)...\sigma_z^{(j)}\mathbf{W}(t_2)\sigma_z^{(\ell)}\mathbf{W}(t_1)|{ \mathbf{i}}\rangle \nonumber  \label{gencorr}
\end{eqnarray}
where it is implied that
\begin{eqnarray}
\sigma_z^{(\kappa)} = \underset{1}{\sigma_o}\otimes\underset{2}{\sigma_o}...%
\underset{\kappa-1}{\sigma_o}\otimes\underset{\kappa}{\sigma_z}\otimes%
\underset{\kappa+1}{\sigma_o} ... \otimes\underset{N}{\sigma_o}  .\label{SoSf}
\end{eqnarray}
Here ${\mathbf{W}}$ and $\sigma_z^{(\kappa)}$ are in the same lexicographically ordered Ising spin basis.
$|\mathbf{i}\rangle $=$|\mathbf{f}\rangle$ are the initial and final state vectors that correspond to the equilibrium distribution such that $\mathbf{W}\ket{\mathbf{i}}=\ket{\mathbf{i}}$. An example is given in the Appendix.


\section{First Spectrum}\label{sec:flux}
\subsection{Flux Noise}
The coherent magnetization of the spins strongly couples to the SQUID's magnetic flux because of the high estimated areal spin density.
In addition, the magnetization noise spectrum can be related to the imaginary part of the susceptibility $P(\omega)=4\chi^{\prime\prime}(\omega)/\beta\omega$, by the fluctuation-dissipation theorem\cite{Reim1986,Vitale1989,McDermott2009}.
If all the surface spins couple to the SQUID equally, the flux noise from the $\ell^{th}$ spin-cluster is
\begin{equation}
P_{\phi }^{(\ell)}(\omega )=2\mu _{o}^{2}\mu _{B}^{2}\frac{\rho }{\pi }%
\frac{R}{r}\displaystyle\int_{0}^{\infty }\displaystyle\sum_{i,j=1}^{N}\langle {s_{i}(0)s_{j}(t)}\rangle e^{\imath\omega t}dt~~
\end{equation}%
where $R$ is the radius of the loop, $r$ is the radius of the wire, $R/r=10$ (see Ref.[\onlinecite{McDermott2009}]) and $\rho$ is the surface spin density.

Using Eq.\ref{gencorr}, $\sum\langle{s_{i}(0)s_{j}(t)}\rangle=\langle\mathbf{f}| \sigma^{i}\mathbf{W}(t)\sigma^{j}\mathbf{W}(0)|\mathbf{i}\rangle$ is calculated by considering all possible combinations of two-point autocorrelation functions ($i=j$) and cross-correlation ($i\neq j$) functions for a given cluster. Here $\mathbf{W}$ is a $2^{N_j}\times2^{N_j}$ flipping probability matrix. Each cluster is assumed to be sufficiently far apart and noninteracting and the total flux noise power is
\begin{equation}
P_{\phi}(\omega)= \displaystyle\sum_\ell{P^{(\ell)}_{\phi}(\omega)}
\end{equation}%

At finite temperatures, for two interacting spins in the $\gamma_j=1$ limit, it can be analytically shown (see Appendix) that the correlation functions are:
\begin{equation}
\langle{s_i(0)s_j(t)}\rangle= \frac{1}{2}e^{-2\Gamma'_-|t|} + \left(\delta_{ij}-\frac{1}{2}\right)e^{4\beta J}e^{-2\Gamma'_+|t|}
\label{Tpt-G}
\end{equation}
where $\Gamma'_{\pm}=[1+\exp(\pm2\beta J)]^{-1}$. Note that $\sum_{ij}\langle{s_i(0)s_j(t)}\rangle= 2e^{-2\Gamma'_-|t|}$. In this case the two interacting TLS behave like a single quasi-spin with effective flipping rate $\Gamma'_-$.

The sum of all \textit{two-point} correlation functions for an arbitrary number of interacting spins can always be expressed as
\begin{eqnarray}
\displaystyle\sum_{i,j}\langle{s_{i}(0)s_{j}(t)}\rangle=\displaystyle\sum_\nu {C_{\nu }e^{-2\Gamma _{\nu }t}}.
\label{CorrFit}
\end{eqnarray}
where, from the master equation it follows that $\Gamma_\nu$s are eigenvalues of $\mathbf{V}$.

\begin{figure}
\centering
\includegraphics[width=1\columnwidth]{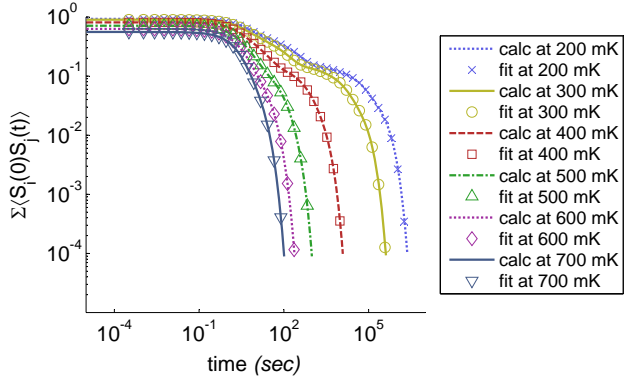}
\caption{ {\bf(a)} Comparison between calculated net correlation functions and fits to $\sum C_\nu\exp(-\Gamma_\nu t)$ at various temperatures. The fits are in perfect agreement with the calculations done for 50 spin clusters, each with 7 spins.}
\label{fig:fits}
\end{figure}

This can also be used to numerically fit to $\sum_{ij}\langle{s_{i}(0)s_{j}(t)}\rangle$ for all the clusters.
In fig.\ref{fig:fits}, the fits are shown to be in exceptionally good agreement with the calculations. A total of 7 terms were used in the expansion over $\nu$ in Eq.\ref{CorrFit}.
Fitting the net correlation functions to Eq.\ref{CorrFit} first, is advantageous for numerics.
In the case of ferromagnetic interactions, the correlations functions can be very long lived at low temperatures. This can be a significant problem for numerically calculating the Fourier transform for the power spectrum.
Instead the fitting parameters (to $\sum C_{\nu }e^{-2\Gamma _{\nu }t}$) can be used to directly obtain the power spectrum, which will just be a sum of Lorentzians weighted by $C_\nu$.

\begin{figure}
\centering
\includegraphics[width=1\columnwidth]{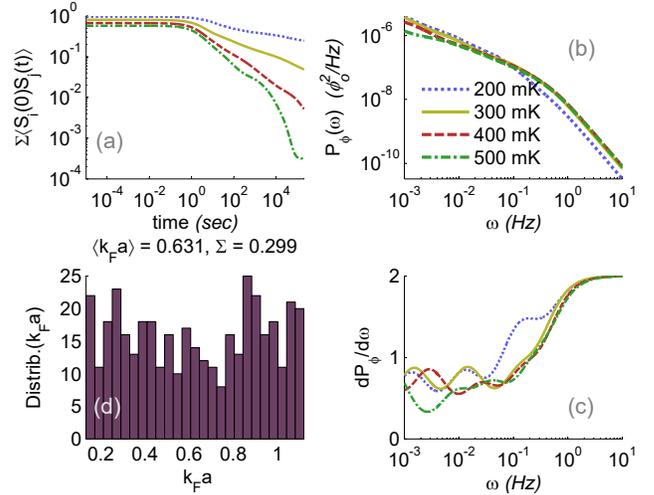}
\caption{ {\bf(a)} Net correlation function for 400 spin clusters, each with 6-9 spins. {\bf(b)} The corresponding power-spectrum for the flux noise, $P_\phi$ and {\bf(c)} the slope of the flux noise.  {\bf(d)} Shows the distribution of $k_Fa$ for each spin cluster. The mean and standard deviation ($\Sigma$) are as indicated.}
\label{fig:P1}
\end{figure}

The temperature dependent net correlation function, power-spectrum of the flux noise, $P_\phi(\omega)$, and its respective slope is shown in Figs.\ref{fig:P1}(a)-(c). For these calculations 75 spin clusters were considered where each cluster has a random $k_Fa$ and between 6-9 spins. The normalized lattice constants ($k_Fa$) were uniformly distributed as shown in Fig.\ref{fig:P1}-(d).
If Eq.\ref{J_RKKY} is expanded upto second order for small $k_Fa$, then it can be shown that:
 \begin{equation}
 J_{ij}\propto\frac{1}{k_FR_{ij}}
 \end{equation}
This leading order term implies that a uniform distribution of the lattice constants will result in $\sim1/J_{ij}$ distribution of interaction strengths, which in turn gives $\sim 1/f$ noise.

\begin{figure}
\centering
\includegraphics[width=1\columnwidth]{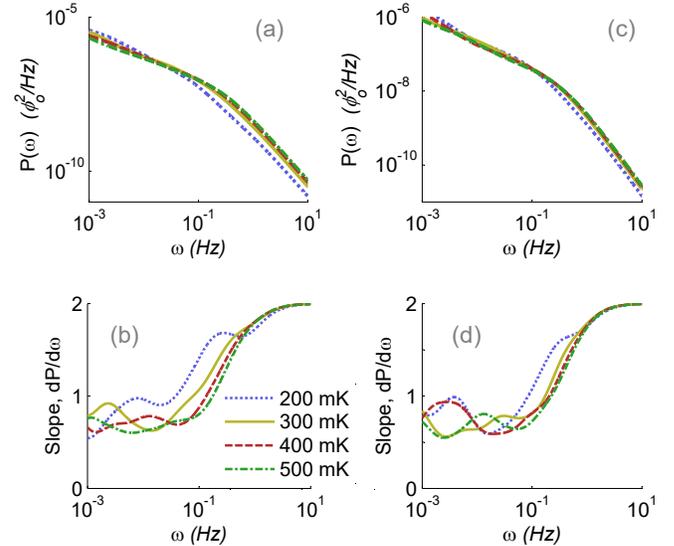}
\caption{
Power spectrum comparison for different cluster sizes. {\bf(a)} Power spectrum and {\bf(b)} corresponding noise slope for 75 cluster with 6-9 spins. {\bf(c)} Power spectrum and {\bf(d)} corresponding noise slope for 20 clusters with 6 spins.}
\label{fig:P1_2}
\end{figure}

From the experimentally estimated areal surface spin density\cite{Sendelbach2008,Bluhm2009} of $\rho\sim5\times10^{17}~m^{-2}$, one can estimate $k_F$ and the average spin separation $\langle{a}\rangle$. Note that these results are independent of the cluster size or the number of spins in a cluster. In Fig.\ref{fig:P1_2}, the noise spectra calculations are repeated for 20 clusters, with 6 spins per cluster.
It can be seen that the noise spectra in Fig.\ref{fig:P1_2} is qualitatively very similar to that of Fig.\ref{fig:P1}. This is consistent with experiments where the flux noise was seen to be more or less independent of the area of the SQUID\cite{Koch1983,McDermott2009} or the cluster size.

As shown in Figs.\ref{fig:P1} (c) and \ref{fig:P1_2} (c), at very high frequencies, the $1/f^{\alpha}$ flux noise power spectra has a slope of $2$ which corresponds to the Lorentzian tail of the noise power. For an intermediate range of frequencies, a region of slope ($\alpha<\sim 1$) is seen at the highest temperature. Eventually for all temperatures, $\alpha\rightarrow 0$ at very low frequencies.

Here the $1/f^{\alpha}$ noise (with $\alpha<1$ like the experiments\cite{Sendelbach2008}) is shown to manifest naturally from the combination of long range ferromagnetic interactions and multiple clusters with a normal distribution of lattice constants.
In the infinite temperature limit or if the interactions are turned off for these calculations, then the same distribution of $k_Fa$ results in just a simple Lorentzian power spectra.
Finally, the upper and lower frequency cutoffs for the $1/f$ type behavior depends on the distribution of $k_Fa$ and the interaction strength -- which is also evident from the temperature dependence in Figs.\ref{fig:P1} and \ref{fig:P1_2}.

\begin{figure}
\centering
\includegraphics[width=1\columnwidth]{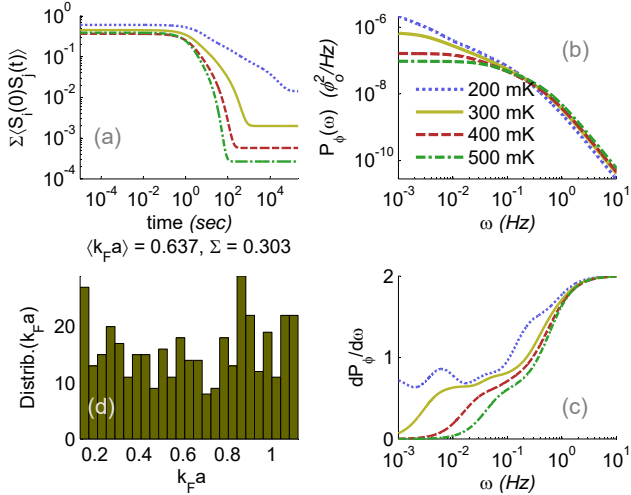}
\caption{ Calculations for spin-clusters with nearest neighbor ferromagnetic interactions showing the {\bf(a)} net correlation function. {\bf(b)} the corresponding power-spectrum for the flux noise and {\bf(c)} the slope of the flux noise.  {\bf(d)} Shows the distribution of $k_Fa$ for each spin cluster. The mean and standard deviation ($\Sigma$) are as indicated. The calculations are for 400 clusters with 6-9 spins.}
\label{fig:P1nni}
\end{figure}


To get a better understanding of the role of the interactions and the cluster-size randomness, we next consider a NNI model with ferromagnetic interactions for each cluster.
The ferromagnetic interaction strength was of the order of $J_o$ and depends on the distance between nearest neighbors and hence varies randomly.
The NNI model also results in a $1/{f^{\alpha}}$ noise spectrum at 200 and 300 mK, as shown in Fig.\ref{fig:P1nni}.

This is similar to the case of the RKKY interactions.
Even though the RKKY interactions oscillate between being ferromagnetic to being antiferromagnetic - the neighboring spins will always experience a large ferromagnetic interaction (if $J_o<1$ and for small $k_Fa$).

However, the variations with temperature are larger for the case of the NNI model as finite size effects are more pronounced for the small spin clusters considered here. For the NNI, $1/f$ noise type behavior is obtained here only for one temperature value, whereas experimentally this behavior was observed over a wide range of temperatures with almost no variation\cite{Koch1983,Sendelbach2008,McDermott2009}. Hence the ferromagnetic NNI \emph{small-spin cluster} model can be ruled out due to the weaker nature of the $1/f$ noise effects mentioned above.


\subsection{Noise Exponent $\alpha$:}\label{sec:alf}
Experimentally, $\alpha<1$ for $1/f^{\alpha}$ flux noise\cite{Koch1983,Sendelbach2008,McDermott2009,Kumar2016pra}. The Ising-Glauber dynamics model self-consistently gives $\alpha<1$ without assuming any distribution of fluctuation rates. A normal distribution of $k_Fa$ is assumed however, which when combined with RKKY interactions gives $\alpha<1$.

The noise slope is obtained from the flux noise spectra in the earlier section
\begin{eqnarray}
\alpha(T,\omega)=\frac{d{P_\phi}}{{d\omega}},
\end{eqnarray}
The frequency dependence can be integrated out within a finite spectral window where $\alpha\sim 1$ at low temperatures,
\begin{equation}{
\alpha(T)=\frac{1}{\omega_u-\omega_l}\displaystyle\int_{\omega_l}^{\omega_u}\alpha(T,\omega) d\omega.
\label{alf}
}
\end{equation}
This range of temperatures over which $\alpha$ remains relatively constant depends on the range of frequencies over which $\alpha(T,\omega)$ is integrated. Here a $10^{-3}$ to $10^{-1}$ Hz frequency window was chosen based on where the $1/f$ noise was predominant.

As shown later in fig.\ref{fig:C3pt_alf} and fig.\ref{fig:CvX}, there is a range of T over which $\alpha$ does not vary much.
This model explains some of the temperature dependence of $\alpha$ seen in flux noise experiments\cite{Koch1983,Sendelbach2008,McDermott2009}.
In some experiments flux noise was almost independent of temperature \cite{Koch1983,Sendelbach2008,McDermott2009}.
However in some of the early experiments\cite{Wellstood1987}, it was not so under {\it all} circumstances. While there was no temperature dependence bellow 1K, there was a low temperature dependence for certain parameters and materials, such as for PbIn/Nb and Pb/Nb\cite{Wellstood1987}. Quite strikingly, for the same set of used materials (i.e. PbIn/Nb for the SQUID's loop/electrode), the flux noise can be either independent- or inversely-proportional or even oscillatory with temperature depending on the construction.
Such conflicting and some times opposite temperature dependence of $1/f$ noise is not uncommon for glassy systems\cite{Massey1997,McCammon2002}.

Much of this oscillatory $\alpha(T)$ behavior arises when the frequency dependence is retained in $\alpha$.
In figs \ref{fig:P1} (c) and \ref{fig:P1_2} (c) it can be seen that for any particular frequency slice, $\alpha$ is not completely constant and shows some sort of oscillatory type behavior as a function of temperature, which is similar to Ref:-\onlinecite{Wellstood1987}. This variation though is small.
This sort of behavior is expected for the RKKY interactions considered here, especially if the lattice constant is small.
Low temperatures more drastically affect $\alpha$ and also make it more oscillatory.

\section{Phase transitions, Three Point Correlations and Flux-Inductance Noise Cross Spectrum}\label{sec:cross}

In this section the time reversal symmetry breaking phase transitions are closely examined.
This is done using three point correlation functions which directly related to the  inductance-flux noise cross spectrum\cite{Sendelbach2008}.
The results are compared with $\alpha$.

It is shown that $1/{f^\alpha}$ noise occurs in a glassy phase where each cluster acts like a macro-spin.
The experiments will not directly observe any time reversal symmetry breaking phase transitions but instead observe the cross-spectrum noise.
The three point correlations further signify that the same mechanism produces both the flux noise and inductance noise. The SQUID's surface spins show a net polarization in the experiments\cite{Sendelbach2008} as the $1/{f^\alpha}$ inductance noise was found to be highly correlated with the $1/{f^\alpha}$ flux noise.

\subsection{Three Point Correlations}\label{sec:C3pt}

This relation between the inductance- and flux noise relates to the three point correlation functions defined here as:
\begin{eqnarray}
C_{3pt} \equiv \frac{1}{N^3}\displaystyle\sum_{i,j,k}\langle s_{i}(0)s_{j}(t_{1})s_{k}(t_{2})\rangle
\label{C3pt}
\end{eqnarray}
The following expression gives the flux- and inductance noise cross power spectrum
%
\begin{widetext}
\begin{eqnarray}\label{Pw3pt}
P_{L\phi}(\omega )=\frac{1}{k_{B}T}\left( 2\rho \mu _{o}^{2}\mu_{B}^{2}\frac{R}{r}\right)^{\frac{3}{2}}\times
\int\limits_{\omega _{a}}^{\omega _{b}}\iint\limits_{0}^{~~~~\infty }
C_{3pt}(t_0,t_1,t_2)
e^{\imath\omega _{-}\tau }e^{\imath\omega _{+}\tau ^{\prime }}d\tau d\tau ^{\prime
}d\omega^{\prime}
\end{eqnarray}%
\end{widetext}
where, $\tau =t_{1}-t_{0}$, $\tau ^{\prime }=t_{2}-t_{1}$, $\omega _{\pm}=\omega \pm \omega ^{\prime }$ and $\omega _{b}-\omega _{a}$ defines the
bandwidth.

\begin{figure}
\centering
\includegraphics[width=0.9\columnwidth]{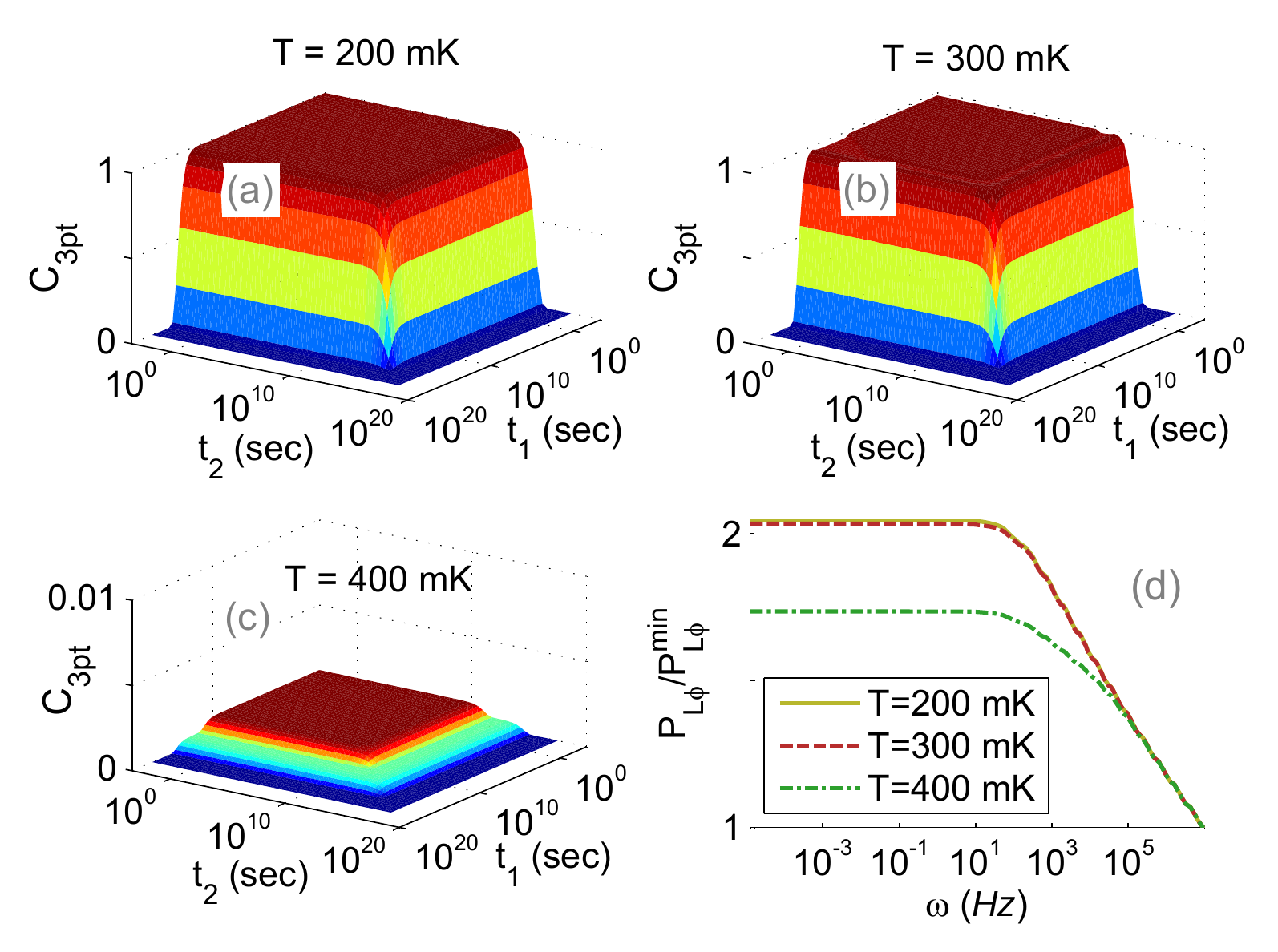}
\caption{ Sum of all three-point auto-correlation and cross-correlation functions, {$C_{3pt}s$}, for $N=9$ spins with ferromagnetic RKKY interactions at ({\bf a}) $T=200~mK$ ({\bf b}) $T=300~mK$ and ({\bf c}) $T=400~mK$ where note that the z-axis is $10^{3}$ times smaller. ({\bf d}) shows the corresponding normalized power spectrum.}
\label{fig_3pt}
\end{figure}

In the experiments $P_{L\phi}$ was found to be inversely proportional to T and $\sim1$ roughly below $100mK$.
$P_{L\phi}$ depends on the sum of all $C_{3pt}$s. As inductance is even under time inversion and magnetic flux is odd, the flux-inductance $C_{3pt}$ can only be nonzero if time reversal symmetry is broken such that there is some \emph{net magnetization}($\mathcal M$) in the sample.
This indicates the appearance of long range ferromagnetic-type-order.
Though antiferromagnetic interactions break time-reversal symmetry, they give negligibly small $C_{3pt}$, since there is no net magnetization as a macroscopic observable.
In terms of bounds without an external magnetic field, $C_{3pt} \leq \mathcal{M} /N^3$. For ferrimagnetic effects (which are present in this system), $C_{3pt} < \mathcal{M} /N^3$.

$C_{3pt}\sim \mathcal{O}(10^{-7})$ for RKKY interactions with antiferromagnetic $J_o$ (for these small spin clusters at initial times) because $\mathcal{M}$ is very small.
This in turn happens because for the range of $kFa$s, the anti-ferromagnetic part dominates in the RKKY interaction. On a regular bipartite lattice this then leads to the magnetization of the two sub-lattices cancelling each other.

Therefore all of this and the flux noise features discussed earlier, indicates that $J_o$ must be ferromagnetic.
To understand  $P_{L\phi}$, the $C_{3pt}$ (for all spin combinations) is calculated by for a single cluster of $9$ spins with ferromagnetic RKKY interactions. The $C_{3pt}$s are extremely long lived at the two lower temperatures.
At initial times, $C_{3pt}\sim 1$ at 200 mK and 300 mK but decreases by over two orders of magnitude at 400 mK (see Fig.\ref{fig_3pt}).
In fig.\ref{fig_3pt}-(d), the power spectrum for these $C_{3pt}$s is normalized for easier comparison since the raw power spectrum for 400 mK is much smaller.



\subsection{Magnetic Phase Transitions}\label{sec:Ph}


In this section, we show superparamagnetic phase transitions occur for these small clusters by examining $C_{3pt}(T)$ and showing that the relaxation rates scale as per an Arrhenius law. At temperatures much lower than the Cuire temperature, transitions from the ferromagnetic to superparamagnetic phase are known to occur in small single-domain nanomagnets\cite{Dormann1988,Knobel2008,Petracic2010}.
This is very similar to the case here, where due to the small magnetic cluster size, the thermal energy becomes comparable to the magnetic anisotropy energy per particle (which is what holds the net magnetic moment). When this happens the cluster will transition from a ferromagnetic state to a superparamagnetic state at the blocking temperature ($T_f$).
This is also similar to the crossover from a thermally activated regime to a quantum tunneling regime\cite{Oelsner2013}.

\begin{figure}
\centering
\includegraphics[width=1\columnwidth]{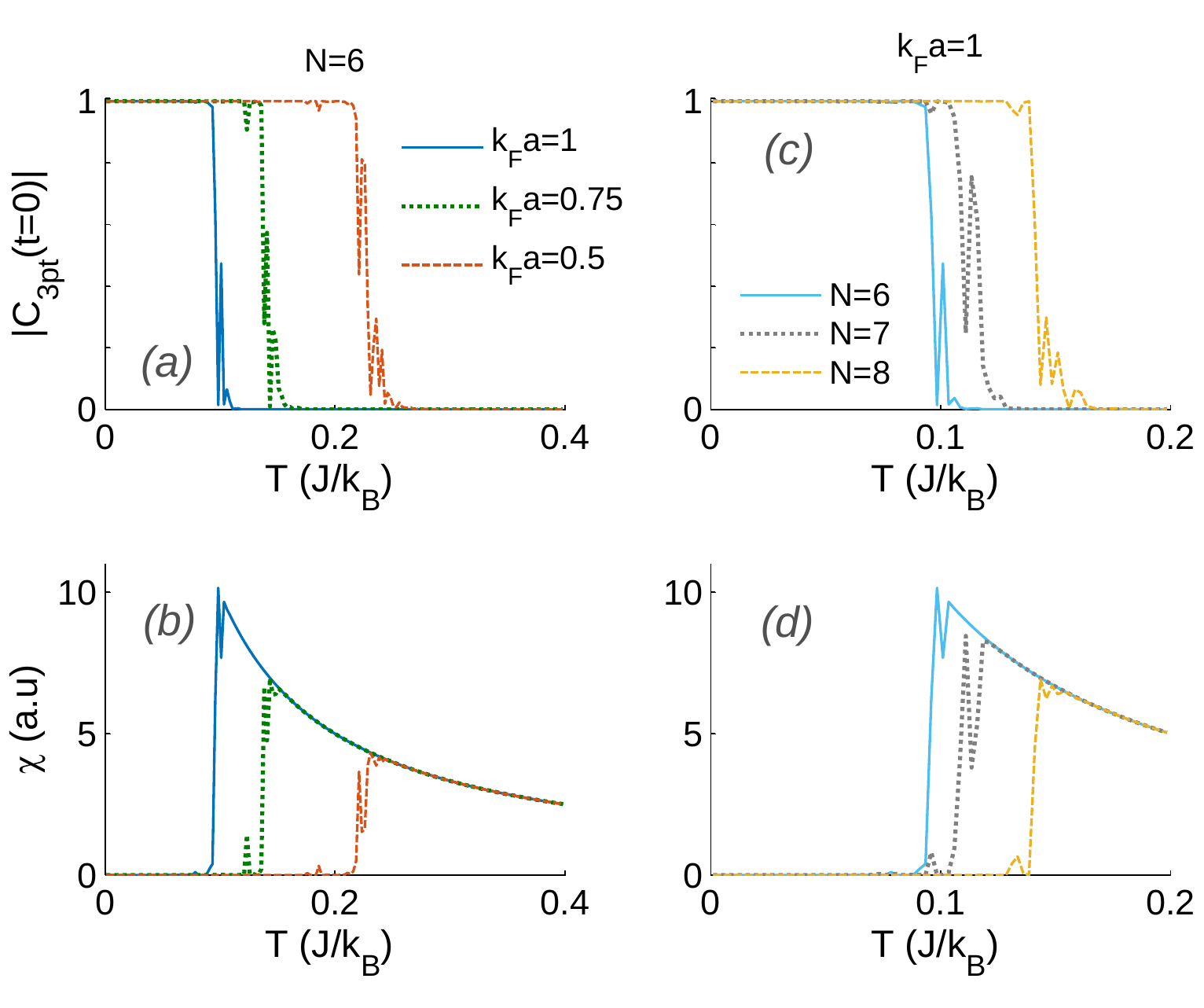}
\caption{ Superparamagnetic phase transitions for a single TLS-cluster as a function of temperature.
({\bf a}) Sum of three-point correlations, $C_{3pt}$ and ({\bf b}) susceptibility, $\chi$ shown for $N=6$ with varying $k_Fa$.
and ({\bf c}) $C_{3pt}$ and ({\bf d}) $\chi$ for varying $N$ with $k_Fa=1$.
All values are obtained at initial times ($t_1=t_2=0$). }
\label{fig:C3pt}
\end{figure}

We examine the temperature dependence of $C_{3pt}$ and ignore its time dependence since $C_{3pt}$ is extremely long lived ($\sim 10^{15}$ sec), and mostly flat and symmetric in $t_1$ and $t_2$.
Hence it is sufficient to consider $C_{3pt}$ only at $t_1=t_2=0$.
These values are shown in fig.\ref{fig:C3pt} as a function of normalized $T$ for different $k_Fa$ and $N$.

\begin{figure}
\centering
\includegraphics[width=1\columnwidth]{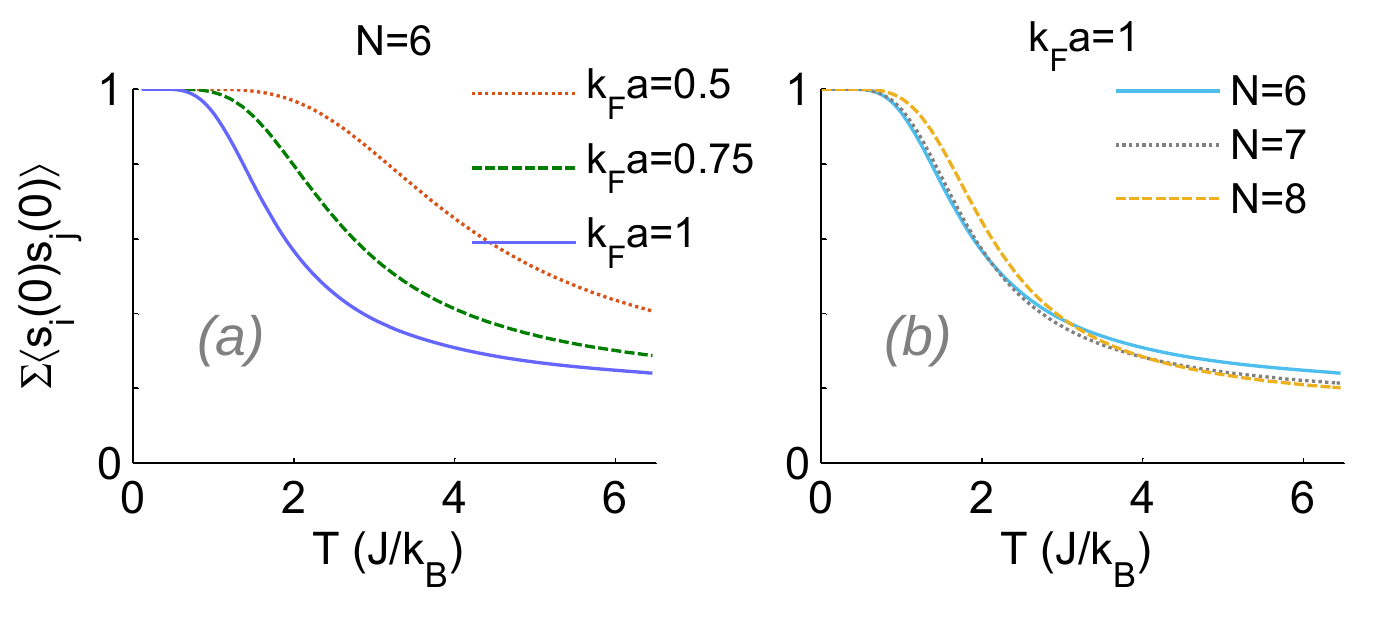}
\caption{ Sum of two-point correlations as a function of normalized temperature at initial times ($t=0$) shown for ({\bf a}) $N=6$ with varying $k_Fa$ and ({\bf b}) varying the number of spins with $k_Fa=1$.}
\label{fig:C2pt}
\end{figure}

Fig.\ref{fig:C3pt} clearly shows very distinct \emph{magnetic phase transitions}.
The critical temperature $T_f$ (or blocking temperature) inversely depends on $k_Fa$ as expected, since smaller $k_Fa$s lead to stronger interactions. In addition the $C_{3pt}$ persists longer with $T$ as $N$ increases.
Here the phase boundaries are sharp even though the cluster size is small. These are sudden superparamagnetic phase transitions which occur because of the small size as explained below. Whereas in case of the two-point-correlations (see fig.\ref{fig:C2pt}), the transitions at the Curie temperature are gradual crossovers as expected far from the thermodynamic size limit.

In addition the susceptibility (at $t=0$)is also shown in fig.\ref{fig:C3pt}.
\begin{eqnarray}
\chi \propto \frac{1}{T}  \displaystyle\sum_{ij}\left( \langle s_i s_j\rangle - \langle s_i\rangle \langle s_j\rangle \right)
\label{X}
\end{eqnarray}
The phase transitions also appear in the order parameter $\chi$ which follows $C_{3pt}$.
This is because $N^2\sum\langle s_j\rangle=C_{3pt}$ everywhere, in this temperature range, except at $T_f$ (where the difference is negligible and around $\mathcal{O}(10^{-8})$).
%
Note that $\chi$ in fig.\ref{fig:C3pt} was calculated for equal time correlations, $t_1=t_2=0$. If $t_2$ is increased then the peak position of $\chi$ in fig.\ref{fig:C3pt} does not change {\emph at all}, as a function of $T$, for an extremely long period of time. For e.g., for N=6, $T(\chi_{max})$  does not change until $t_2 > 10^{14}$ sec.

For a single cluster the phase transitions at $T_f$ are sharp looking for the $C_{3pt}$ order parameter.
However over a small range near $T_f$, several very sharp fluctuations occur as shown in fig.\ref{fig:C3pt}.
This is because the thermal energy near $T_f$ is similar to the magnetic anisotropy energy.
If a $dT$ were finer, more $1\leftrightarrow 0$ fluctuations for $C_{3pt}$ would appear as a function of $T$.
This is a key result, because it implies that small temperature fluctuations can cause rapid fluctuations in $M$. This behavior indicates that each cluster acts like a flip-flopping macro-spin (with magnetic moment $N\mu_B$) or a source of random-telegraphic-magnetic-noise.


The magnetic RTN behavior can be better understood by examining the system dynamics near crticality.
The $T$-dependent scaling laws for the relaxation times can be characterized\cite{Dormann1988,Reich1990prb,Mori2003prb,Wu2011prb} as the system approaches $T_f$.
%
%
A single cluster's overall spin relaxation time $\tau_r$, was obtained by fitting
$\sum \langle{s_i(0)}{s_j(t)}\rangle/N^2$ to $\mathcal{A}\exp(-t/\tau_r)$, where the curve-fits were excellent.
Next, $\tau_r$ is further fit to a simple Arrhenius equation
\begin{eqnarray}
\tau_r=\tau_0\exp(E_b/T)
\end{eqnarray}
where $E_b$ is the barrier energy and $\tau_0$ is a constant.
For some non-linear systems, $\tau_0$ can also be temperature dependent\cite{Malakhov2002}.

\begin{figure}
\centering
\includegraphics[width=1\columnwidth]{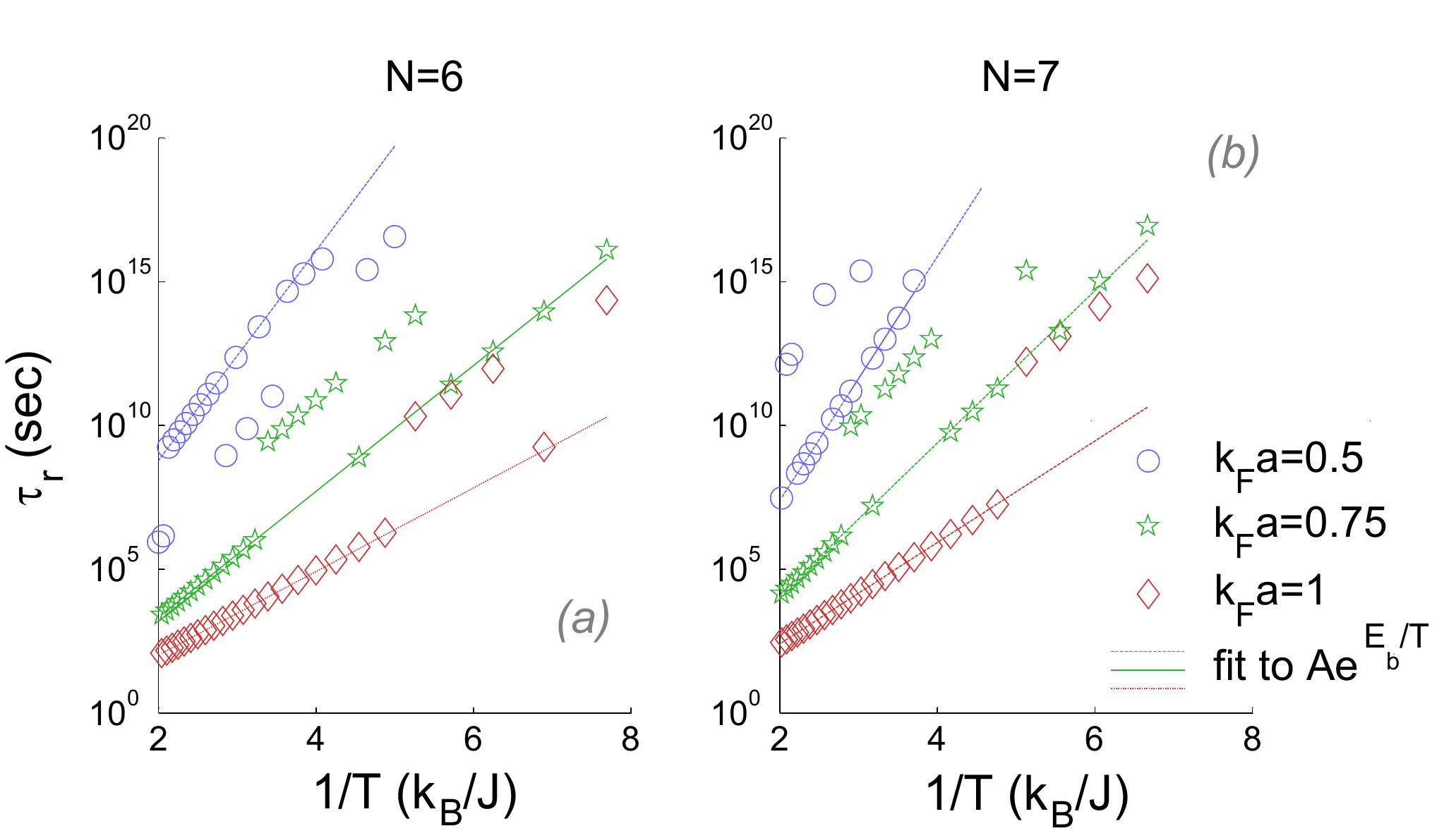}
\caption{ Relaxation times $\tau_r$ as a function of inverse temperature, leading up to $T_f$, shown for (a) N=6 (b) N=7 for various $k_Fa=1$. The lines are fits to an Arrhenius equation $\tau_0\exp(E_b/T)$. }
\label{fig:TauAhr}
\end{figure}

In fig.\ref{fig:TauAhr}, $\tau_r$ is shown as a function of $1/T$ leading up to $T_f$ for N=6 and 7 for various $k_Fa$s.
Note that there are two distinct relaxation time scales for each $k_Fa$ in these small clusters.
The $\tau_r$ exponentially slows down and follows an Arrhenius law as shown in Fig.\ref{fig:TauAhr}. On a logarithmic scale, the simple Arrhenius form $\tau_r$s are straight lines as a function of $1/T$. Small deviations from the straight lines are shown as $T$ approaches $T_f$ and as $k_Fa$ gets smaller.
For $k_F=1$, $\tau_o=[0.11,0.07,0.04]$ sec and the barrier energy $E_b=[3.42,4.09,5.26]~J/k_B$ for N = 6,7 and 8 respectively.
These parameters are reasonable for the small system size. As expected $E_b\propto T_f$ and increases as N increases.
Overall, the Arrhenius scaling of $\tau_r$ indicates SPTs.

In in fig.\ref{fig:TauAhr}, for all the cases, a prominent second relaxation time scale appears as $T\rightarrow T_f$.
This is due to another local energy minima with its own spin configuration -- an indication of glassy behavior.
The macro-spin relaxation rate is very different from the main time scale, even though they have similar energies.
Small temperature fluctuations can lead to the selection of either state.
The second set of $\tau_r$s are higher, their scaling is less steep than the Arrhenius law and slightly deviate from it as $k_Fa$ gets smaller.
The second time scale is more prominent and well separated for smaller N, and as N increases it merges with the first time scale.

The sum of two-point correlations (at $t=0$) should also be examined and is shown in fig.\ref{fig:C2pt} as a function of T. Strong finite size effects are seen. At large T, the two-point-correlations gradually decay to $=1/N$ (sum of \emph{T-independent} auto-correlations). As $N$ increases the transitions become sharper and tend to 0.
Above $T_f$, $M=0$, but the two-point correlation function is not zero (compare fig.\ref{fig:C3pt} and fig.\ref{fig:C2pt}) and hence this denotes a glassy paramagnetic phase.

\subsection{$1/f^{\alpha}$ Noise from an Ensemble of Magnetic Phase Transitions}\label{sec:MC}

\begin{figure}
\centering
\includegraphics[width=0.8\columnwidth]{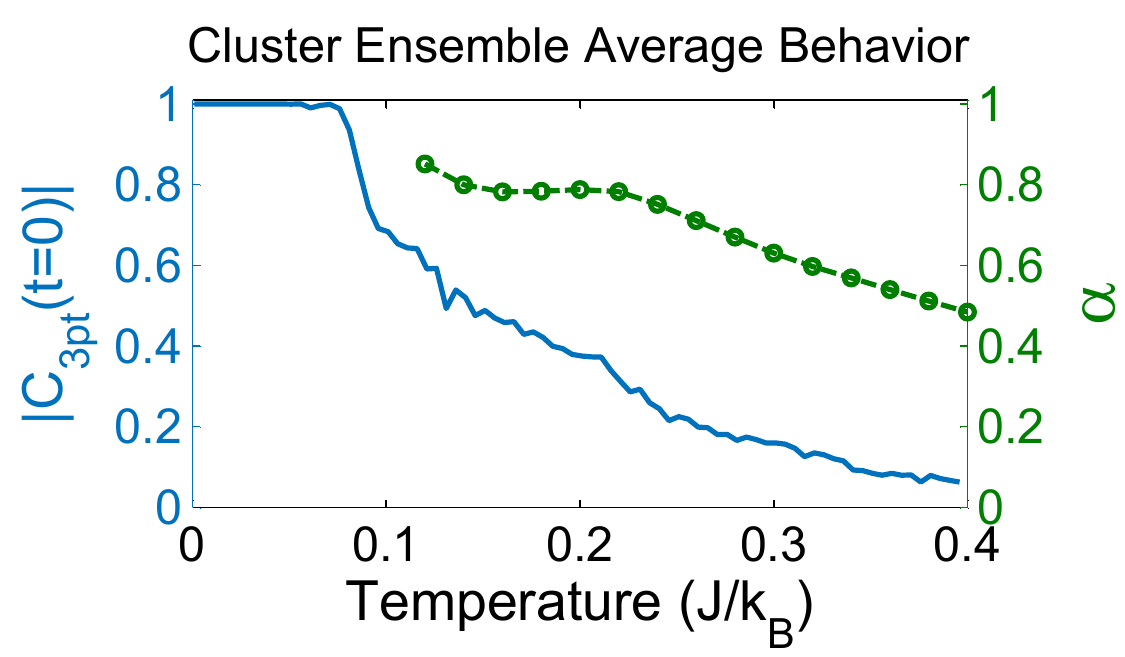}
\caption{ Left axis shows the ensemble averaged sum of three-point correlations ($C_{3pt}$) as a function of temperature (at $t_1=t_2=0$ initial time). The ensemble average is over the \emph{exact} same 75 random $k_Fa$ and $N$, which gives the noise exponent, $\alpha$ on the right-axis.}
\label{fig:C3pt_alf}
\end{figure}

In the previous section it was shown how a single cluster becomes a source of magnetic RTN because of SPTs.
In order to better understand the cause of the $1/f^{\alpha}$ noise, we next compare the cluster ensemble averaged $C_{3pt}$ with $\alpha$.
Fig.\ref{fig:C3pt_alf} shows the ensemble averaged $C_{3pt}$ and $\alpha$, as a function of normalized T.
The ensemble average was over the \emph{exact} same 75 random $k_Fa$ and $N$, which gave the $1/f^\alpha$ noise in fig.\ref{fig:P1_2}-(a).

Remarkably, $1/f^{\alpha}$ noise occurs in the same temperature range where the ensemble averaged $C_{3pt}$ gradually transitions to 0.
This is the temperature range in which each cluster in the superparamagnetic phase can act like a macro-spin.
There is a range of temperatures over which $\alpha$ does not vary much, which is based on the $10^{-3}~to~10^{-1}$ Hz frequency integration window.
Here $\alpha\sim 0.8$ when more clusters are in the ferromagnetic phase. $\alpha(T)$ falls off as fewer clusters remain ordered.

It should also be pointed out that the experiments would not directly observe the phase-transitions of fig.\ref{fig:C3pt} because
it is not easy to isolate a single cluster.
Instead, because of the ensemble averaging effects, a gradual temperature dependent decline would be seen.
This decline in  Fig.\ref{fig:C3pt_alf} is further augmented by a $1/T$ factor in Eq.\ref{Pw3pt} for the flux-inductance cross spectrum noise\cite{Sendelbach2008}.


\subsection{$1/f^{\alpha}$ Noise and Crossovers: Monte Carlo Simulations}\label{sec:MC}

Since it is unlikely that the experiments can directly observe the magnetic phase-transitions of fig.\ref{fig:C3pt} for a single cluster -- it is important to find other distinct features that might still be experimentally observable despite the ensemble averaging.
In addition to the SPTs discussed in sec.\ref{C3pt}, crossovers from the superparamagnetic to paramagnetic phase (or pseudo phase-transitions) are also expected to occur at higher Curie temperatures.
For small clusters far from the thermodynamic limit, the specific heat $C_v$ has the appearance of a smooth crossover rather than a sharp phase-transition.
These crossovers would be experimentally observable and would provide a further validation of this cluster model.
Also because of finite size effects, the critical temperature, $T'_c$, at which these crossovers occur can provide more insight into the number of spins in a cluster and their interaction strengths.

In this section a numerical Monte-Carlo method is used to compliment the main correlation function technique used in this paper.
More specifically, the temperature dependent phase transitions were obtained using a Monte Carlo technique with parallel-tempering. The order parameters was the specific heat: $C_v=(\avg{E}^2-\avg{E^2})/T^2$, where $E$ is the total energy.
For the Monte-Carlo steps, first a complete sweep of all the TLS in all lattices was taken at various temperatures.
Here unlike NNI models, since RKKY interactions are involved, while updating a spin configuration, the acceptance decision is made based on the energy of the entire interacting cluster. This is numerically feasible for all systems sizes considered here.
Next for the parallel tempering part, the different replicas at different temperatures was swapped using the detailed balance condition: $min\left(\mathcal{R},\exp[(E_1-E_2)(T_1^{-1}-T_2^{-1})/k_B]\right)$, where $0<\mathcal{R}<1$ is random.
The whole process was repeated over $10^4$ times.

The $T$ dependent $C_v$ is also compared with $\alpha$.
Since the following comparisons between pseudo phase-transitions and $\alpha$ involved two different calculation methods, a number of steps were taken to ensure that the calculations were fully consistent. Exactly the same spin-cluster configuration and seeds for random-number-generation was used to generate the $k_Fa$ distribution for both methods.
Both Monte-Carlo and $\alpha(T)$ calculations were done for 75 spin clusters , each with 6-9 spins and a uniform distribution was chosen for $k_Fa$.

\begin{figure}
\centering
\includegraphics[width=1\columnwidth]{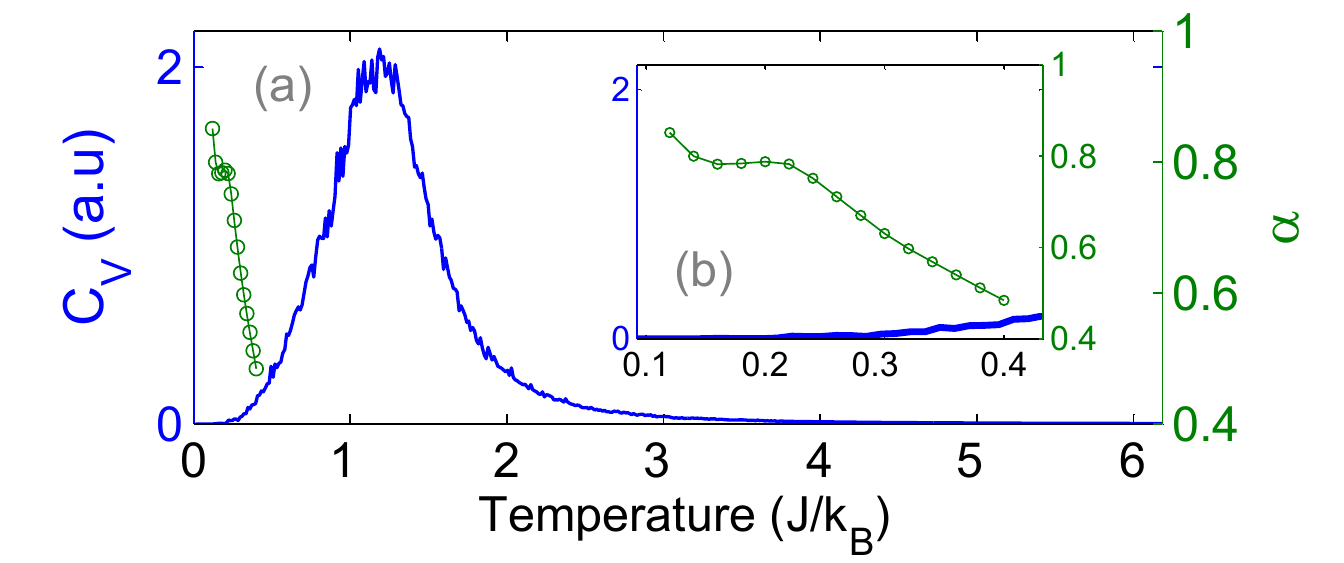}
\caption{ (a) Comparison of the noise exponent $\alpha$, against the specific heat, $C_v$ for 75 spin clusters, each with 6-9 spins. $C_v$ is calculated using the Monte-Carlo method. 
This is the same distribution of spin-clusters that gives the $1/f$ noise shown in fig.\ref{fig:P1_2}-(a) with $\langle k_Fa\rangle=0.63$.
(b) A close-up of $\alpha$.}
\label{fig:CvX}
\end{figure}

$C_v$ is shown in fig.\ref{fig:CvX} along with $\alpha$ as a function of temperature. A close up view of $\alpha$ over a narrower range of temperatures is shown in the figure inset.
A second-order pseudo phase-transition is shown in fig.\ref{fig:CvX}-(a) which occurs at $T'_c\approx 1.2 Jk_B^{-1}$. 
At the lowest temperature $\alpha$ tends to rise towards $1$.
As discussed earlier, at very low T as the system heads towards $T_f$ it is difficult to extract $\alpha$ since a single cluster's correlations times scale exponentially as per an Arrhenius law.

\begin{figure}
\centering
\includegraphics[width=0.85\columnwidth]{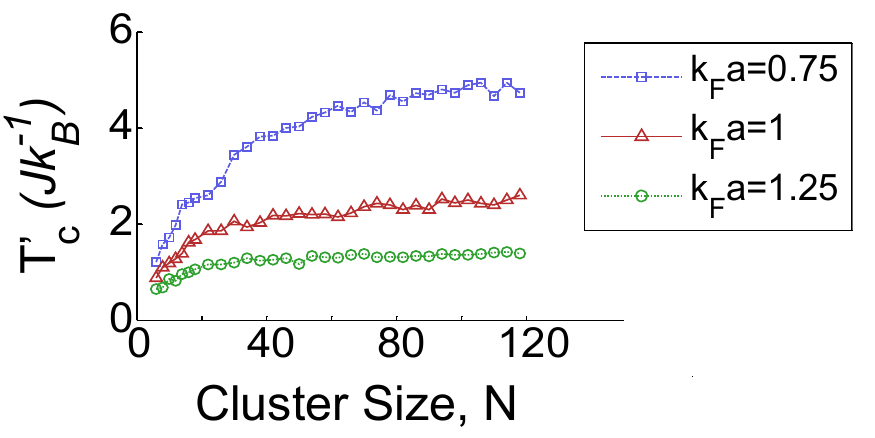}
\caption{Critical temperature, $T'_c$, from Monte Carlo simulations for the spin clusters with RKKY interactions as a function of cluster size, $N$ for various normalized lattice spacing $k_Fa$. }
\label{fig:TcN}
\end{figure}

There are strong finite size effects in this system.
For clusters with $6-9$ spins the transition takes place at a lower $T'_c$, diverging away from the thermodynamic limit.
Compare this to $T'_c\sim2.2~Jk_B^{-1}$ for a disorder free large-N limit 2D NNI model.
Fig.\ref{fig:TcN} shows $T'_c$ as a function of $N$, for a single cluster with ferromagnetic RKKY interactions, for various $k_Fa$. These were obtained from Monte Carlo simulations.
Because of the RKKY interactions, $T'_c$ is very sensitive to $k_Fa$ and $N$ for small sized systems.
An experimental measurement of $T'_c$, will provide more insight into the average number of spins in a cluster and their interaction strengths.

Overall it is shown in this section that the distinct phase-transition like peaks would still be observable for $C_v$ crossovers despite the cluster ensemble averaging. This crossovers occur at temperatures that are well above where $1/f^{\alpha}$ noise occurs. An experimental observation of this will validate this model.

\section{Second Spectrum: Inductance Noise}\label{sec:Ind}
This section discusses various features of the inductance noise, $P_L$ which is the associated noise spectrum or the second spectrum. It is a quantitative measure of the spectral wandering of the first spectrum and is interpreted as the
noise of the noise \cite{Nguyen2001}.
In the experiments of Ref.\onlinecite{Sendelbach2008}, $P_L$ was measured bellow $2K$ and varied considerably with temperature.
Despite having the same noise microscopics, it is not obvious why the second spectrum should have this strong temperature dependence while the flux noise (first spectrum) does not vary with temperature. It is analytically shown here why this is the case.

The flux noise is related to the imaginary part of the susceptibility via the
fluctuation-dissipation theorem
\begin{eqnarray}
P(\omega)= 2\hbar\coth\left(\frac{\hbar\omega}{k_BT}\right)\chi^{\prime\prime} = \lim_{k_BT\gg\hbar\omega} 2\frac{k_BT}{\omega}\chi^{\prime\prime}.
\label{FDT}
\end{eqnarray}
Assuming all spins couple to the SQUID equally\cite{McDermott2009,Zhou2010}, the imaginary part of the inductance then relates to the spin susceptibility within a layer of thickness $d$ on the surface as follows
\begin{eqnarray}
L^{\prime\prime} = \mu_o d\frac{R}{r}\chi^{\prime\prime}
\label{LX}
\end{eqnarray}
where $L^{\prime\prime}$ is the imaginary part of the inductance. If $\rho$ is the surface spin density then $d=\rho/\tilde{n}$, where $\tilde n$ is the spin density.
Therefore again from the fluctuation dissipation theorem,
\begin{equation}
\chi ^{\prime \prime }(\omega )=2\frac{\tilde{n}\mu _{o}\mu _{B}^{2}\omega }{%
k_{B}T}\displaystyle\sum_{i,j}\displaystyle\int_{0}^{\infty }\langle
s_{i}(0)s_{j}(t)\rangle e^{\imath\omega t}dt.
\end{equation}%

The sum of all two-point correlation functions for the system of interacting spins can be expressed in terms of the eigenvalues of $\mathbf{V}$ as follows,
\begin{eqnarray}
\displaystyle\sum_{i,j}\langle{s_{i}(0)s_{j}(t)}\rangle=\displaystyle\sum_\nu {C_{\nu }e^{-2\Gamma _{\nu }t}}.
\label{2ptcorr}
\end{eqnarray}
Also see the two spin example in Eq.\ref{A.Tpt}.
%
Hence
\begin{equation}
\chi ^{\prime \prime }(\omega )=2\tilde{n}\mu _{o}\mu _{B}^{2}\frac{\omega }{%
k_{B}T}\sum_{\nu }\frac{C_{\nu }\Gamma _{\nu }}{\Gamma _{\nu
}^{2}+\omega ^{2}}
\end{equation}%
and the real part from Kramers-Kronig relation is
\begin{eqnarray}
\chi^{\prime}(\omega)&=&\frac{2}{\pi }{\mathcal{P}}\displaystyle\int_{0}^{\infty }\frac{\chi ^{\prime\prime }(\omega^{\prime} )}{\omega ^{\prime 2}-\omega ^{2}}d\omega ^{\prime }\\
&=&\frac{2\tilde{n}\mu_{o}\mu_{B}^{2}}{k_{B}T}\sum_{\nu }\frac{C_{\nu }\Gamma _{\nu }^{2}}{\Gamma _{\nu }^{2}+\omega ^{2}}~.
\end{eqnarray}%
where, $\mathcal{P}$ is Cauchy's principal value. Hence from the total susceptibility $\chi(\omega)=\chi^{\prime}(\omega)+i\chi ^{\prime\prime}(\omega)$:
\begin{eqnarray}
\chi (t)=\int_{0}^{\infty }\chi(\omega)e^{\imath\omega t}d\omega
=\frac{2\tilde{n}\mu _{o}\mu _{B}^{2}}{k_{B}T}\sum_{i,j}\langle
s_{i}(0)s_{j}(t)\rangle.~~~
\label{Xt}
\end{eqnarray}
The real part $\chi^\prime$ has to be considered in order to establish the $T^{-1}$ dependence analytically.

The inductance noise can be generally expressed as
\begin{eqnarray}
P_{L}(\omega ) 
&=&\left( \mu _{o}d\frac{R}{r}\right) ^{2}\int_{0}^{\infty }\langle \chi
(0)\chi (t)\rangle e^{\imath\omega t}dt  \label{PL}
\end{eqnarray}%
The inductance noise can then be further explicitly expressed in terms of the
spectral density of the dynamical four-point noise correlation functions\cite{Kogan.book},
\begin{eqnarray}
P^{[2]}_L(\omega) = \left(2\rho\frac{\mu_o^2\mu_B^2}{k_BT} \frac{R}{r}%
\right)^2 \displaystyle\iint\limits_{\omega_{a}}^{~~~~%
\omega_{b}}S^{[2]}(\omega,\omega_1,\omega_2)d\omega_1 d\omega_2~~~~~~
\label{PL2}
\end{eqnarray}
where
\begin{widetext}
\begin{equation}
S^{[2]}(\omega,\omega_1,\omega_2) =  \iiint\limits_0^{~~~~\infty} \displaystyle
\sum_{j,k,l,m}\langle s_j(t_1)s_k(t_2)s_l(t_3)s_m(t_4)\rangle  e^{i(\omega_a-\omega)\tau'}e^{i(\omega_b+\omega)\tau'' }e^{i\omega\tau}
d\tau^{\prime} d\tau^{\prime\prime} d\tau
\label{S2i}
\end{equation}
\end{widetext}
here $\tau^{\prime }=t_2-t_1$, $\tau^{\prime \prime }=t_4-t_3$ and $\tau=t_4+t_3-t_2-t_1$, $\Delta\omega=\omega_{b}-\omega_{a}$ is the
bandwidth within which the second spectrum is observed and $j,k,l,m$ are spin indices.

Numerical second spectrum calculations can be considerably difficult. The Gaussian approximation for the four-point correlation functions can make these calculations much more feasible.
For the model considered here, to an excellent approximation:
\begin{eqnarray}
\displaystyle\sum_{j,k,l,m}\langle{s_j(t_1)s_k(t_2)s_l(t_3)s_m(t_4)}\rangle~~~~~~~~~~~~~~~~~~\\\nonumber
~~~~~~~~~~~\approx\sum_{j,k,l,m}\langle{s_j(t_1)s_k(t_2)}\rangle\langle{s_l(t_3)s_m(t_4)}\rangle.
\label{Fpt-2pt}
\end{eqnarray}
In Appendix-B, it is shown that this is an \emph{exceptionally} good approximation for ferromagnetic RKKY interactions particularly in the ordered phase and in the temperature regime of interest for $1/f$ noise.
This because the terms in the Gaussian approximation relate to the order parameter (see Appendix-B).
However, note that this is not a good approximation for antiferromagnetic RKKY interactions.


Therefore, using the Gaussian approximation
\begin{equation}
S^{[2]}(\omega,\omega_1,\omega_2) =  \delta(\tau)P^{\prime}(\omega,\omega_a)P^{\prime}(\omega,\omega_b)
\label{S2ii}
\end{equation}
where $P'(\omega,\omega_{a(b)})=\sum C_{\nu}[2\Gamma_{\nu} + i (\omega-\omega_{a(b)}) ]^{-1}$
based on the two-point correlations.
Substituting this into Eq.\ref{PL2}, the following expression is obtained for the associated spectrum of correlated Ising spin fluctuations.
\begin{widetext}
\begin{eqnarray}
P^{[2]}_L(\omega) &\approx&  \delta(\tau)\left(2\rho\frac{\mu_o^2\mu_B^2}{k_BT} \frac{R}{r}\right)^2   \left|\left( \sum\limits_{\nu}C_{\nu}
\log\left[\frac{\omega_{b} + \omega + 2i\Gamma_{\nu}}{\omega_{a} + \omega + 2i\Gamma_{\nu}}\right]\right)
\left(\sum\limits_{\nu}C_{\nu}
\log\left[\frac{\omega_{b} - \omega + 2i\Gamma_{\nu}}{\omega_{a} - \omega + 2i\Gamma_{\nu}}\right]\right)
\right|
\label{P2_int}
\end{eqnarray}
\end{widetext}
$P^{[2]}_L$ looks similar to a Lorentzian spectral function where its amplitude is determined by $\Delta\nu=\omega_{b}-\omega_{a}$.

\begin{figure}
\centering
\includegraphics[width=1\columnwidth]{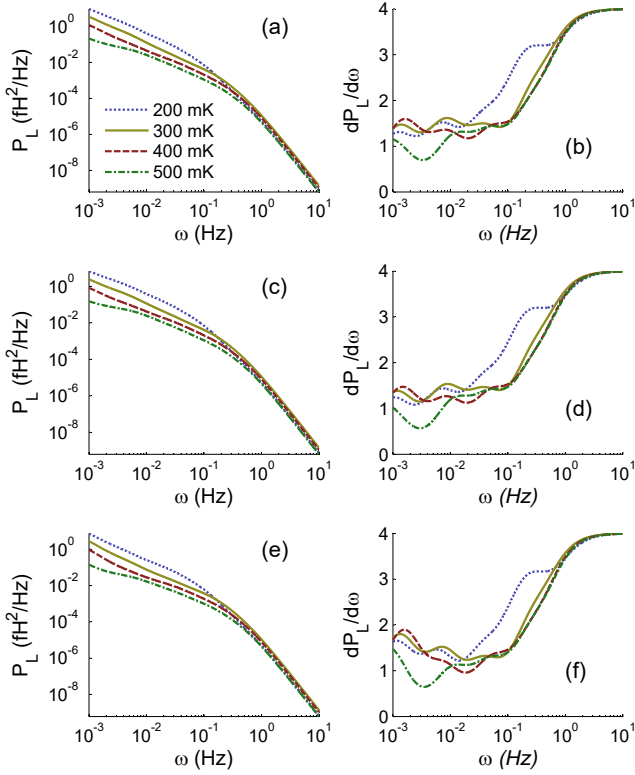}
\caption{Inductance noise plots for different $\Delta\nu=\omega_b-\omega_a$ showing the \textbf{(a)} Power-spectrum and \textbf{(b)} its respective slope for $\omega_b=10$ Hz and $\omega_a=10^{-3}$ Hz.
  \textbf{(c)} Power-spectrum and \textbf{(d)} its respective slope for $\omega_b=10$ Hz and $\omega_a=10^{-2}$ Hz.
 \textbf{(e)} Power-spectrum and \textbf{(f)} its respective slope for $\omega_b=10$ Hz and $\omega_a=10^{-1}$ Hz. Calculations are for 400 cluster with 6-9 spins.}
\label{fig:P2}
\end{figure}

Eq.\ref{P2_int} is used for the numerical inductance noise calculations for the ensemble of spin-clusters.
The T-dependent inductance noise spectrum and its slope is shown in fig.\ref {fig:P2} for different $\Delta\nu=\omega_b-\omega_a$, since second spectrum calculations are sensitive to $\Delta\nu$.
In subplots fig.\ref {fig:P2}-(a) and (b) nearly the whole spectrum is covered. While in the other subplots the $\omega_a$ is consecutively reduced by an order of magnitude. Varying $\omega_a$ affects the power spectrum more than varying $\omega_b$. It is shown that when $\Delta\nu$ nearly covers the full spectrum, the noise power spectrum now shows $1/f^{\alpha }$ behavior at intermediate frequencies. The average integrated $\alpha$ between $0.001-0.05~Hz$ varies from $\sim 1.57$ (at $200~mK$) to $\sim1.24$ (at $500~mK$). As the temperature further increases $\alpha\rightarrow 0$. The $\alpha=4$ at high frequencies is due to the square of the Lorentzian tail while at the lowest frequencies $\alpha$ eventually rolls over to zero. Overall the inductance noise shows a large variation with temperature and the calculations agree very well with experiment\cite{Sendelbach2008,Wellstood2011}.

\section{summary}
In this paper, a self-consistent model is proposed that explains various observed features of the temperature dependent $1/f^{\alpha(T)}$ magnetization noise in SQUIDs.
$1/f^\alpha$ flux noise with $\alpha\sim 0.8$ over a range of low temperatures and frequencies. The flux noise is further shown to be independent of system size.
The model comprises of multiple finite sized spin clusters with ferromagnetic RKKY interactions.
Similar noise results can be obtained with other long range ferromagnetic interactions.
This model is representative of spatially disorder from defects or substitutional impurities.
Furthermore, here both a random normal distribution of cluster sizes and ferromagnetic long range interactions are essential for obtaining $1/f^\alpha$ noise.
There is no a priori assumption made on a log normal distribution of fluctuation rates for obtaining the $1/f^\alpha$ noise.
All results, including $\alpha\sim 0.8$, are obtained self-consistently from Glauber dynamics.


Calculated cross-correlations between flux- and inductance-noise with ferromagnetic RKKY interactions shows that there is a magnetically ordered phase for the TLS, as seen in experiment\cite{Sendelbach2008}.
The cross spectrum is obtained from three-point correlation functions.

Furthermore, the three-point correlation and susceptibility order parameters reveal sharp-looking low temperature superparamagnetic phase transitions. This happens in the temperature range where $1/f^\alpha$ magnetization noise is seen.
Each cluster transitions from a ferromagnetically ordered phase to a glassy paramagnetic phase.
Recent experiments suggest a dynamical paramagnetic environment due to the asymmetry in the flux noise for positive and negative frequencies\cite{Quintana2017}.

A single cluster behaves like a macro-spin in this SPT phase transition, where even the smallest temperature fluctuations near criticality can lead to random telegraphic net-magnetization fluctuations.
However the experiments would not directly see the phase-transitions of fig.\ref{fig:C3pt} because of the ensemble averaging effects.
Here $1/f^{\alpha}$ noise is the observable for an ensemble of clusters that are individually phase transitioning.

Monte Carlo specific heat calculations also show pseudo phase transitions from the superparamagnetic to paramagnetic phase
at the higher Curie temperature.
On the temperature scale, $1/f^\alpha$ noise occurs prior to the onset of the pseudo phase transition.
This would be experimentally observable despite ensemble averaging and would further validate this model.
Overall as a function of temperature, there appear to be atleast four macrostates available to each cluster, one ferromagnetic phase, two superparamagnetic phases and one paramagnetic phase.
Note that each cluster can behave like a macro-spin and that there is some randomness in their positions and the net magnetic moment. If the interactions between the macro-spins then become sufficiently strong, then this can lead to some additional phases\cite{Edwards1975} as the number of clusters approach the thermodynamic limit.

For the inductance noise, based on the fluctuation dissipation theorem, it is analytically shown why the inductance noise is inherently ${\rm T}^{-2}$ dependent while the flux noise is not.
Analytical expressions are provided for easier four-point correlation function calculations.

Finally, the method suggested here for obtaining the $n-$point correlation functions is key to these calculations and is discussed in detail. This is a fully self-consistent method and model, that takes time dependence, spatial correlations, interactions and temperature dependence into account.
By incorporating Glauber dynamics, this method also expands the scope of the existing $1/f$ noise calculation methods in a self-consistent way.
This method is useful for numerics and analytics as shown here.

%



\section{Appendix}\label{App}

\subsection{Example: Two Correlated TLS}\label{sec.A1}
An analytical example is given here for the correlation functions, and the power spectrum for a pair of interacting spins using this paper's model.

\emph{Master Equations:}
If the average occupation of the two states is the same (for unbiased fluctuators), then for single spin-$i$ $\mathbf{V}_i = \gamma_i(\sigma_x-\sigma_o)$, where $\gamma_i$ is the $i^{th}$ spin's relaxation rate.
$\gamma_i$ is a phenomenological parameter. It will be shown that the Ising-Glauber model naturally gives a temperature dependent relaxation rate from the spin-flip probability.

The corresponding flipping probability matrix is
\begin{equation}
\mathbf{W}_i(t)=\exp(\mathbf{V}_it)=\frac{1+e^{-2\gamma_it}}{2}\sigma_o+\frac{1-e^{-2\gamma_i t}}{2}\sigma_x.
\label{A.W}
\end{equation}
For $N$ uncorrelated (non-interacting) Ising spins, it is straightforward to express the entire system's flipping probability matrix as a tensor product of individual $\mathbf{W}_i$s:
\begin{eqnarray}
\mathbf{W}=\mathbf{W}_1\otimes\mathbf{W}_2\otimes...\mathbf{W}_N
\end{eqnarray}
The matrix $\mathbf{V}=\dot{\mathbf{W}}\mathbf{W}^{-1}$  which is also
\begin{eqnarray}
\mathbf{V} = -\displaystyle\sum_j\gamma_j{\bf{I}} + \displaystyle\sum_j\gamma_j\sigma_x^{(j)}
\end{eqnarray}
where $\bf{I}$ is the identity matrix. For self consistency it can be verified that this is the same as what is obtained from Eq.\ref{Glaub} in the $T\rightarrow\infty$ limit.

\emph{Correlated TLS Fluctuations:}
In order to obtain the power spectrum of a given pair of interacting TLSs or Ising spins, the corresponding two point correlation functions have to be obtained first. Consider the Ising-Glauber model outlined in the main body of the paper for two fluctuating Ising spins that are correlated. If $B=0$, then the $\mathbf{V}$ matrix in the $\{11,1\bar1,\bar11,\bar1\bar1\}$ basis has the following form

\begin{eqnarray}
\mathbf{V}&=&\footnotesize{\frac{2}{\varepsilon}\left[
\begin{array}{cccc}
  -\displaystyle\sum_i\gamma_i&   \gamma_2 \varepsilon'               & \gamma_1\varepsilon'                &               0\\
           \gamma_2     & -\varepsilon'\displaystyle\sum_i\gamma_i    &                   0                      & \gamma_1\\
           \gamma_1     &                   0                      & -\varepsilon'\displaystyle\sum_i\gamma_i    & \gamma_2\\
                0       &   \gamma_1\varepsilon'                & \gamma_2 \varepsilon'                 & -\displaystyle\sum_i\gamma_i
\end{array}
\right]}~~~~
\label{A.V_2CoRTS}
\end{eqnarray}
or
\begin{eqnarray}
\mathbf{V} &=& \mathbf{V}_o\times\left[ I + \tanh(\beta{J}) \sigma_z^{(1)}\sigma_z^{(2)}\right]
\label{A.V2int}
\end{eqnarray}
where $\varepsilon' = \exp(2\beta J )$ and $\varepsilon = 1+\exp(2\beta J)$ and
\begin{eqnarray}
\mathbf{V}_o = \gamma_1\sigma_x^{(1)} + \gamma_2\sigma_x^{(2)}- (\gamma_1+\gamma_2){\bf{I}}
\label{A.V}
\end{eqnarray}
Obviously if $\beta J=0$, then $\mathbf{V}=\mathbf{V}_o$ which is just the $\mathbf{V}$ matrix for two uncorrelated TLS. Also $\sigma_x^{(1)}=\sigma_x\otimes\sigma_o$ and $\sigma_x^{(2)}=\sigma_o\otimes\sigma_x$.

\emph{Limits:}
More general expressions for the correlation functions and the power spectrum are given in Ref.[\cite{De2015noise}].
In the $\displaystyle{T\rightarrow\infty}$ limit all cross-correlations are zero and the two-point autocorrelation function just reduces to that of two uncorrelated Ising spins/TLS:
\begin{eqnarray}
\langle s_i(0)s_j(t)\rangle = \delta_{ij}e^{-2\gamma_it}
\label{A.ucorr2}
\end{eqnarray}

Whereas at finite temperatures and in the $\gamma_j=1$ limit,
the flipping probability matrix ($\mathbf{W}=\exp(-\mathbf{V}t)$) has the following form:
\begin{widetext}
\begin{eqnarray}
\mathbf{W} = \frac{1}{2\varepsilon}\left[
\begin{array}{cccc}
\varepsilon'e^{-4t} + \varepsilon e^{-4t\varepsilon'/\varepsilon} + 1, &              1 - e^{-4t}, &     1 - e^{-4t}, &  \varepsilon' e^{-4t} - \varepsilon e^{-4t\varepsilon'/\varepsilon} + 1\\
-\varepsilon'(e^{-4t} - 1), &  e^{-4t} + \varepsilon(1+e^{-4t/\varepsilon}) - 1, &  e^{-4t} + \varepsilon(1-e^{-4t/\varepsilon}) - 1, &  -\varepsilon'(e^{-4t} - 1) \\
-\varepsilon'(e^{-4t} - 1), &  e^{-4t} + \varepsilon(1-e^{-4t/\varepsilon}) - 1, &  e^{-4t} + \varepsilon(1+e^{-4t/\varepsilon}) - 1, &  -\varepsilon'(e^{-4t} - 1) \\
\varepsilon'e^{-4t} - \varepsilon e^{-4t\varepsilon'/\varepsilon} + 1,&               1 - e^{-4t}, &      1 - e^{-4t}, &  \varepsilon'e^{-4t} + \varepsilon e^{-4t\varepsilon'/\varepsilon} + 1
\end{array}
\right]
\end{eqnarray}
\end{widetext}
This gives the corresponding correlation functions for two interacting TLS:
\begin{equation}
\langle{s_i(0)s_j(t)}\rangle= \frac{1}{2}e^{-2\Gamma'_-|t|} + \left(\delta_{ij}-\frac{1}{2}\right)e^{4\beta J}e^{-2\Gamma'_+|t|}
\label{A.Tpt}
\end{equation}
where $\Gamma'_{\pm}=[1+\exp(\pm2\beta J)]^{-1}$. Note that here, $\displaystyle\sum_{ij}\langle{s_i(0)s_j(t)}\rangle= 2e^{-2\Gamma'_-|t|}$. This implies that in this special case the two interacting TLS can be expressed as a single effective TLS with flipping rate $\Gamma'_-$.


\subsection{The Gaussian Approximation for the Four-Point Correlation Function}\label{sec.A2}

\begin{figure}
\centering
\includegraphics[width=0.9\columnwidth]{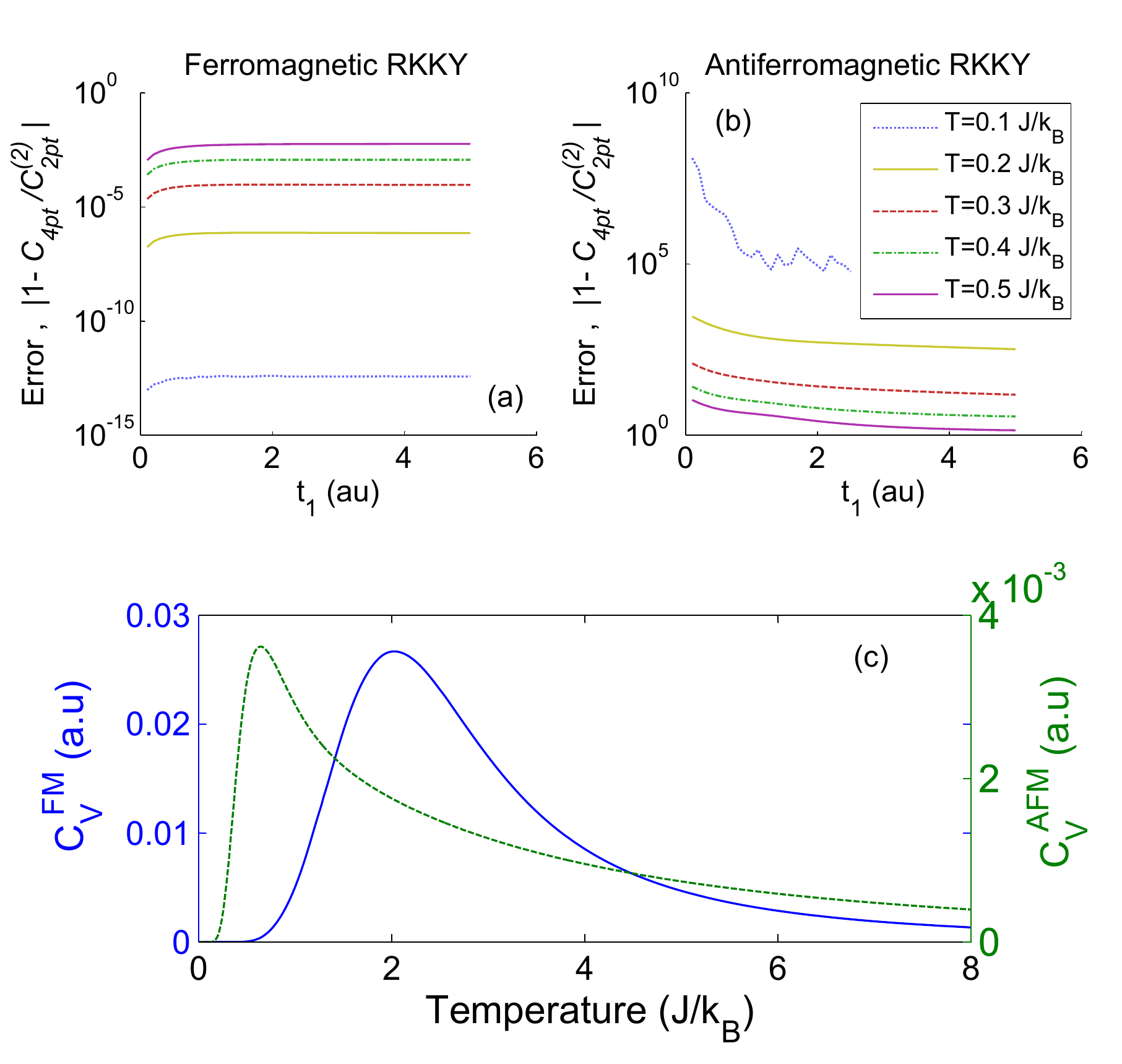}
\caption{ The errors signifying how good the Gaussian approximation is for {\bf(a)} Ferromagnetic(FM) RKKY interaction and {\bf(b)} Antiferromagnetic(AFM) RKKY interactions. Here $t_1=t_2=t_3$ for N=6 and $k_Fa=0.75$.
The results are of the same order for larger N and randomly sampled $t_1$ and $t_2$.
(c) Corresponding specific heat for $t_1=0$.}
\label{fig:CoErr}
\end{figure}

In this appendix, the accuracy of the Gaussian approximation (see Eq. \ref{Fpt-2pt}) is more closely examined. The four-point correlation function and its approximation using the two-point correlations can be defined as:
\begin{eqnarray}
C_{4pt}\equiv\displaystyle\sum_{j,k,l,m}\langle{s_j(0)s_k(t_1)s_l(t_2)s_m(t_3)}\rangle \\
C^{(2)}_{2pt}\equiv\sum_{j,k,l,m}\langle{s_j(0)s_k(t_1)}\rangle\langle{s_l(t_2)s_m(t_3)}\rangle.
\label{A.Fpt-2pt}
\end{eqnarray}
In the Gaussian approximation  $C_{4pt}\approx C^{(2)}_{2pt}$. The accuracy of this can be quantified by the relative error, $1-C_{4pt}/C^{(2)}_{2pt}$ which is shown in Fig.\ref{fig:CoErr}-(a) and (b). In this example $N=6$ and $t_1=t_2=t_3$. Very similar results, where the errors were of the same order of magnitude were obtained for $N=7~\&~8$ and for randomly chosen $t_2~\&~t_3$.
\\

As shown in Fig.\ref{fig:CoErr}-(a), the Gaussian approximation is exceptionally good for ferromagnetic RKKY interactions in the temperature regime of interest for inductance noise.
Whereas, in Fig.\ref{fig:CoErr}-(b) shows that for antiferromagnetic RKKY interactions, the Gaussian is quite poor, especially at low T.
For the ferromagnetic case, the errors are significantly less than $0.001~\%$, in the $1/f$ noise temperature regime (see fig.\ref{fig:CvX} for comparison).
The errors drop exponentially with decreasing temperature. At the T where $C_{3pt}\sim 1$ (see fig.\ref{fig:C3pt}), the errors drop to $\sim 10^{-9}$ for ferromagnetic RKKY interactions.
Hence the Gaussian noise approximation is excellent for calculating the $1/f$ inductance noise.

The reason for this is as follows. Note that $C_{4pt}-C_{2pt}^{(2)}$ is like an order parameter. Infact, the specific heat $C_v \propto (C_{4pt}-C_{2pt}^{(2)})/T^2$, minus the auto-correlation terms at $t=0$.
This is shown in fig.\ref{fig:CoErr}-(c). The $C_v$ curves are smooth (unlike $C_{3pt}$), since $N$ is small, and represent a pseudo phase-transition. As long as the two-point cross correlations persist, the Gaussian approximation holds good. Fig.\ref{fig:CoErr}-(c) also shows why the Gaussian approximation is poor for antiferromagnetic RKKY interactions in the low temperature regime of interest.

It should be pointed out that while the correlation function calculation method used in this paper can reproduce the crossover phase transition features (such as at $T'_c$ for $C_v$) obtained from Monte Carlo simulations -- the opposite was found to be not true. The Monte Carlo calculations could not capture the glass transition at occurs at $T_f$ for $\chi$.



\section{Acknowledgements}
Many thanks to Robert Joynt for several invaluable discussions and for pioneering the quasi-Hamiltonian open quantum systems method. Very special thanks to Robert McDermott for carefully discussing the  experiments. Also my sincere thanks to Leonid Pryadko for several helpful discussions on Monte Carlo techniques.

\bibliographystyle{apsrev}

\end{document}